\documentclass[structabstract]{aa}
\usepackage{graphicx}
\usepackage{txfonts}
\usepackage{natbib}
\bibpunct{(}{)}{;}{a}{}{,}
\usepackage{color}
\newcommand{\modif}[1]{#1}

\begin{document} 
%
%
%
\title{Reconstruction of the CMB lensing for Planck}

%
\author{L. Perotto\inst{1,2}
        \and
        J. Bobin\inst{3,4}
        \and
        S. Plaszczynski\inst{2}
        \and
        J.-L. Starck\inst{3}
        \and
        A. Lavabre\inst{2}
        }
%
\institute{Laboratoire de Physique Subatomique et de Cosmologie (LPSC), 
CNRS : UMR5821, IN2P3, Universit\'e Joseph Fourier - Grenoble I, Institut Polytechnique de Grenoble, France\\ 
           \email{perotto@lpsc.in2p3.fr}
        \and
           Laboratoire de l'Acc\'el\'erateur Lin\'eaire (LAL), 
CNRS : UMR8607, IN2P3, Universit\'e Paris-Sud, Orsay, France\\
            \email{plaszczy@lal.in2p3.fr,lavabre@lal.in2p3.fr}
        \and
Laboratoire AIM (UMR 7158), CEA/DSM-CNRS-Universit\'e Paris Diderot,   
IRFU, SEDI-SAP, Service d'Astrophysique,  Centre de Saclay, 
F-91191 Gif-Sur-Yvette cedex, France \\
            \email{jstarck@cea.fr}
            \and
           Applied and Computational Mathematics (ACM), California Institute of Technology, 
1200 E.California Bvd, M/C 217-50, PASADENA CA-91125, USA\\
            \email{bobin@acm.caltech.edu}
} 
\date{Received November 12, 2009}

%

\abstract{}{
We prepare real-life Cosmic Microwave Background (CMB) lensing extraction with the forthcoming
Planck satellite data, by studying two systematic effects related to the foregrounds contamination: 
the impact of foreground residuals after a component separation on the lensed CMB map, and of removing a large 
contaminated region of the sky. 
}{
We first use the Generalized Morphological Component Analysis (GMCA) method to perform a component separation 
within a simplified framework which allows a high statistics Monte-Carlo study. For the second systematic, we apply
 a realistic mask on the temperature maps and then, restore them using a recent inpainting technique on the sphere. 
We investigate the reconstruction of the CMB lensing from the resultant maps using a quadratic estimator in the flat 
sky limit and on the full sphere.
}{
We find that the foreground residuals from the GMCA method does not alter
significantly the lensed signal, nor does the mask corrected with the inpainting method, \modif{even in the presence of 
point sources residuals}.
}{}


\maketitle
%

\section{Introduction}

Cosmic Microwave Background (CMB) temperature anisotropies and polarisation
measurements have been one of the key cosmological probes to establish the current cosmological constant
$\Lambda$ and Cold Dark Matter ($\Lambda$CDM) paradigm. Reaching the most precise measurement 
of these observables is 
the main scientific goal of the forthcoming or on-going CMB experiments -- such as the 
\emph{European Spacial Agency} satellite {\sc Planck}\footnote{
\tt{http:$//$www.rssd.esa.int$/$index.php?project$=$PLANCK} \tt{\&page$=$index}} 
\modif{which has been successfully launched on the 14th of May 2009 and has currently begun collecting data.} 

{\sc Planck} is designed to deliver full-sky coverage, low-level noise, high resolution
 temperature and polarisations maps~\citep[see][]{Tauber06,Bluebook2005}. With such high quality observations, it 
becomes doable to extract from the CMB maps cosmological informations beyond the 
angular power spectra (two-points correlations, hereafter APS), by exploiting the measurable 
non-Gaussianities (see e.~g.~\citet{Komatsu2002thesis,Bluebook2005}). 

The weak gravitational lensing is one of the sources of non-Gaussianity affecting the CMB  
after the recombination (see \citet{Lewis2006} for a review). The CMB photons are 
weakly deflected by the gravitational 
potential of the intervening large-scale structures (LSS), which perturbs the Gaussian 
statistic of the CMB anisotropies~\citep{Bernardeau1997, Zaldarriaga2000}.
 Conversely, it becomes possible to reconstruct the underlying gravitational potential by exploiting 
the higher-order correlations induced by the weak lensing in the CMB maps~\citep{Bernardeau1997,
Guzik2000,Takada2001,Hu2001a,Hirata2003f}. 

The relevance of the CMB lensing reconstruction for the Cosmology is twofold. First, for the 
sake of measuring the primordial B-mode of polarisation predicted by the inflationary 
models~\citep{Kamionkowski1997,Seljak1997}, 
the CMB lensing is a major contaminant. It induces a secondary B-mode polarisation signal 
in perturbing the E-mode polarisation pattern~\citep{Zaldarriaga1998}. A lensing 
reconstruction allowing the \emph{delensing} of the CMB maps is required to recover the 
primordial B mode signal~\citep{Knox2002,Seljak2004}. However, CMB lensing is also a powerful 
cosmological probe of the matter distribution integrated from the last scattering surface to us. 
In a near future, it is an unique opportunity to probe the full-sky LSS distribution, with 
a maximum efficiency at redshift around 3, where structures still experience a well 
described linear growth~\citep{Lewis2006}. A lensing 
reconstruction would largely improve the sensitivity of the CMB experiments to the cosmological 
parameter affecting the growth of the LSS, such as neutrino mass or dark energy~\citep{Hu2002j,
 Kaplinghat2003, Lesgourgues2006, Perotto2006}.

Although well-known theoretically~\citep{Blanchard1987}, the CMB lensing has never been directly 
measured. \citet{Smith2007} and \citet{Hirata2008} have found evidence for a detection of the CMB 
lensing in the WMAP data by correlating with several other LSS probes (Luminous Red Galaxies, 
 Quasars and radio sources) at 3.4$\sigma$ and 2.5$\sigma$ level respectively. This situation is 
expected to change with the forthcoming {\sc Planck} data. {\sc Planck} will be the first CMB 
experiment allowing to measure the underlying gravitational potential without requiring 
any external data. However, even with the never before met quality of the {\sc Planck} data,
CMB lensing reconstruction will be challenging. CMB lensing is a very subtle 
secondary effect, affecting the smaller angular scale at the limit of the {\sc Planck} 
resolution in a correlated way over several degrees on the sky. As already quoted, CMB lensing reconstruction is 
based on the induced non-Gaussianities in the CMB maps, in the form of mode
coupling. Consequently, any process resulting in coupling different Fourier moments is a challenging systematic to deal 
with in order to retrieve the lensing signal (see \citet{su2009} for a recent study of 
the impact of instrumental systematics on the CMB lensing reconstruction bias). 
Astrophysical components and other secondary effects might also be a source of non-Gaussianity. These components include~: 
Thermal and Kinetic Sunyaev-Zel'dovich effects (thSZ and kSZ), due to the 
scattering of CMB radiation by electrons within the galaxy clusters~\citep{SZ1970};  foreground emissions, such as 
synchrotron, Bremsstrahlung and dust diffuse galactic emission as well as extragalactic point 
sources. All these components may give a sizable 
contribution to the level of non-Gaussianities in the CMB maps~\citep{Aghanim1999, Argueso2003, Amblard2004, Riquelme2007, Babich2008}.

The impact of most of the aforementioned effects on the CMB lensing analysis with WMAP data
 has been investigated by~\citet{Hirata2008}. The fact that they found a negligible 
contamination level is encouraging. However, such a result could change when one considers the higher resolution, 
better sensitivity maps provided by {\sc Planck}. In \citet{Barreiro06}, the component separation impact on 
non-Gaussianity was studied in the framework of the PLANCK project, but no lensing reconstruction was performed. 
Hence, the impact of these foreground residuals on the CMB lensing reconstruction is still to be studied.  

The overall purpose of the present study is to give an insight to the issues we should deal with before 
undertaking any complete study of the CMB lensing retrieval with {\sc Planck}: what is the impact of the 
foreground residuals on the CMB lensing reconstruction? Will it still be possible to reconstruct the CMB
lensing after a component separation process or will such a process alter the temperature map
statistics? How should we deal with the masking issue?  
Beyond the \emph{detection} of the CMB lensing signal, we tackled the 
\emph{reconstruction} of the underlying projected potential APS. We investigated two issues, the 
impact of a component separation algorithm on the lensing 
reconstruction and the impact of a masked temperature map restoration before applying a deflection estimator.    

Sect.~\ref{sec:formalism}  briefly review the CMB lensing effect and the reconstruction method. 
We  present in Sect.~\ref{sec:patches} an analysis of the impact of one component separation technique, named 
\emph{Generalized Morphological Component Analysis} (GMCA)~\citep{Bobin2008}, which 
is one of the different methods investigated by the PLANCK consortium~\citep{Leach2008}.
In Sect.~\ref{sec:fullsky}, we show how a recent gap filling method (i.e. inpainting process) \citep{Abrial2008} 
may solve the masking problem, which may be one of a the major issue for the CMB lensing retrieval 
since it introduces some misleading correlations between different angular scales in the maps.

%
%
%
%
%
%
%

\section{CMB lensing}
\label{sec:formalism}

In this section, we briefly review the CMB lensing effect and the reconstruction method. 
We introduce the notations used throughout this paper. \\

%
%
%
%

CMB photons geodesic is weakly deflected by the gravitational potential from the last
 scattering surface to us. Observationally, this effect results in a remapping of the CMB 
temperature anisotropies $T = \Delta\Theta/\Theta_{\mathrm{CMB}}$, according to~\citet{Blanchard1987}:
\begin{equation}
\widetilde{T}(\hat{\vec n})= T(\hat{\vec n} + \vec{d}(\hat{\vec n})).
\label{eq:remapping}
\end{equation}
In words, the lensed temperature $\widetilde{T}$ in a given direction of the sky $\hat{\vec n}$ is the 
temperature $T$ one would have seen in the neighboring direction $\hat{\vec n} + 
\vec{d}(\hat{\vec n})$ in the absence of any intervening mass.
 The deflection angle, $\vec{d}(\hat{\vec n})$, is the gradient of the line-of-sight projection of the
 gravitational potential\footnote{A priori, the remapping function should depend not only on a 
convergence field but also on a rotation field, so that the deflection angle is not purely gradient but 
have a rotational contribution. However, \citet{Hirata2003} have shown that 
rotation field effect would be negligible for the next generation of CMB experiments.},
 $\vec{d}(\hat{\vec n})=\vec \nabla \phi(\hat{\vec n})$, where $\phi$
 can be calculated within the Born  approximation, as the integral
 along the line-of-sight of the tridimensional gravitational potential~\citep{Challinor2005}.  

The CMB lensing probes the intervening mass in a broad range of redshifts, 
from $z_{*} = 1090$, at the last scattering surface, to $z=0$, with a maximum efficiency at $z\sim
3$. At such a high redshift, the LSS responsible for
the CMB lensing  (with typical scale of $300h^{-1}Mpc$) still
experience a linear regime of growth. As a result,  the projected potential $\phi$ can be assumed to be a
 Gaussian random field; The consequences of the
 non-linear corrections to $\phi$ are shown to be weak on the CMB lensed 
observables~\citep{Challinor2005}. Thus, this hypothesis holds very well as long as the 
CMB lensing study doesn't aim at measuring a correlation with other LSS probes at lower redshifts.

Besides, in the standard $\Lambda $CDM model, the deflection angles
have a rms of  $\simeq 2.7$ arcmin  and can be correlated over several degrees on the
sky. 
The typical scales of the lensing effects  are small enough for a
convenient analysis within the flat sky approximation. The projected 
potential may be  decomposed on a Fourier basis $\phi(\vec{k})$.
and its statistics is completely defined by:
\begin{equation}
    <\phi(\vec k_1)\phi(\vec k_2)> = (2\pi)^2\delta(\vec k_1+\vec k_2)\, C_{k_1}^{\phi\phi}, 
\end{equation}
where $C_{k_1}^{\phi\phi}$ is the full-sky projected potential APS taken at a multipole 
$l=|\vec k_1|$ and it is related to the deflection APS by~:
\begin{equation}
    C_{k}^{\mathrm{dd}}={k}^2C_{k}^{\phi\phi}
\label{eq:dd2phiphi}
\end{equation}

The lensed CMB temperature APS can be derived from the Fourier transform of 
Eq. (\ref{eq:remapping}) (e. g. as in~\citet{Okamoto2003}). Lensing effect slightly modifies the 
APS of the CMB temperature, weakly smoothing the power at all angular scale 
at the benefit of the smaller angular scales. Deeply in the damping tail, at multipole
 $l\gtrsim 3000$, lensing contribution even dominates over the pure CMB one. However, 
the main observational consequences of the CMB lensing effect lies beyond the APS. 
The remapping induces non-Gaussianities in the CMB temperature field, 
in the form of some correlations between different angular scales. 

Consequently, the two-point correlation function of the lensed temperature modes, calculated at
 the first order in $\phi$, writes~\citep{Okamoto2003}:
\begin{eqnarray}
\langle \widetilde{T}(\vec k_1) \widetilde{T}(\vec k_2) \rangle_{\mathrm{CMB}} \,&=&\, (2\pi)^2\delta(\vec k_1+\vec k_2)\, \widetilde{C}_{k_1}^{\mathrm{\, TT}} \\\nonumber
 &+&\, f_{\mathrm{TT}}(\vec k_1,\vec k_2) \, \phi(\vec L) + O(\phi^2)
\label{eq:correl_lens}
\end{eqnarray}
where $\vec L = \vec k_1 + \vec k_2$, and the $\mathrm{CMB}$ subscript denotes an ensemble average over 
different realisations of the CMB but a fixed integrated potential field. The weighting function $f_{\mathrm{TT}}$ 
depends on the primordial temperature APS, such as:
\begin{equation}
f_{\mathrm{TT}}(\vec k_1,\vec k_2) =\vec L \cdot \vec k_1 \, C_{k_1}^{\mathrm{TT}} + \vec L \cdot \vec k_2 \, C_{k_2}^{\mathrm{TT}}.
\label{eq:fTT}
\end{equation}
Similarly, one can calculate the four-point correlation function of the CMB temperature field 
-- such as in \citet{Kesden2003}. One finds that the trispectrum of the lensed temperature field 
-- or equivalently, the connected part of its four-point correlation function -- is non-null 
even if the underlying (unlensed) temperature field is purely Gaussian. 

%
%
%
%

For the sake of reconstructing the integrated gravitational potential field from a lensed CMB map,
two methods exist in the literature. First, the quadratic estimator approach has been 
developed in~\citet{Hu2001a, Hu2002, Okamoto2003}. Then a maximum-likelihood 
estimator method has been derived in~\citet{Hirata2003f,Hirata2003}.
 The latter approach may increase the capabilities of the highest sensitivity highest resolution CMB
 projects in reconstructing the integrated potential while the former method is still very close to 
the optimality for current built experiment such as {\sc Planck}. Thus we adopt the quadratic estimator 
throughout this work. 

In the flat sky approximation, the estimated potential map takes the following form\citep{Okamoto2003}:
\begin{equation}
\widehat{\phi}_{\mathrm{TT}}({\vec L}) =  \frac{A_{\mathrm{TT}}(L)}{L^2} \int \frac{d^2\vec k_1}{(2\pi)^2}\, 
\bar{T}(\vec k_1) \bar{T}(\vec k_2) F_{\mathrm{TT}}(\vec k_1,\vec k_2)\,,
\label{eq:estimator}
\end{equation} 
where the Fourier modes $\bar{T}(\vec k)$ refer to the \emph{observed} temperature modes, 
affected by both the CMB lensing and the instrumental noise of the CMB experiment concerned. 
More precisely, the temperature map is assumed to be contaminated by an additional white 
Gaussian noise and deconvolved from a beam function assumed to be Gaussian, so that its 
APS reads:
\begin{equation}
\langle \bar{T}(\vec k_1) \bar{T}(\vec k_2) \rangle \,= \, (2\pi)^2\,\delta(\vec k_1+\vec k_2)\, \left(\widetilde{C}_{k_1}^{\mathrm{\, TT}}+ N_{k_1}^{\mathrm{TT}}\right),
\label{eq:obscltt}
\end{equation} 
where $N_{k_1}^{\mathrm{TT}}$ is the instrumental noise APS, modeled in this analysis as:
\begin{equation}
N_{k}^{\mathrm{TT}} \, = \, \theta_{\mathrm{fwhm}}^{\, 2} \sigma_{\mathrm{T}}^2 \, \, exp \left[ k^2 \frac{\theta_{\mathrm{fwhm}}^{\,2}}{8\ln 2}\right],
\label{eq:nltt}
\end{equation} 
where $\theta_{\mathrm{fwhm}}$ is the full-width at half maximum (FWHM) of the beam function and 
$\sigma_{\mathrm{T}}$, the root mean square of the noise per resolution elements.  

Besides, the normalisation function is calculated so that $\widehat \phi_{\mathrm{TT}}({\vec L})$ is an 
unbiased estimator of the integrated potential field:
\begin{equation}
A_{\mathrm{TT}}(L) \,=\,  L^2 \left[ \int \frac{d^2\vec k_1}{(2\pi)^2}\,  f_{\mathrm{TT}}(\vec k_1,\vec k_2)
F_{\mathrm{TT}}(\vec k_1,\vec k_2) \right]^{-1} \,.
\label{eq:att}
\end{equation}  
Then, the weighting function $F_{\mathrm{TT}}$ is adjusted to minimize the dominant contribution to 
the estimator variance, i. e. the unconnected part of the quantity
$\langle \hat{\phi}_{\mathrm{TT}}({\vec L}) \hat{\phi}_{\mathrm{TT}}({\vec L'})\rangle -
(2\pi)^2\, \delta(\vec L+\vec L')C_L^{\phi\phi}$. Derived in
\citet{Okamoto2003}, the calculation leads to:

\begin{equation}
F_{\mathrm{TT}}(\vec k_1,\vec k_2) \,=  \, \frac{f_{\mathrm{TT}}(\vec k_1,\vec k_2)}
{2\bar{C}_{k_1}^{\mathrm{TT}} \bar{C}_{k_2}^{\mathrm{TT}}},
\label{eq:optimal}
\end{equation}
where $\bar{C}_{k}^{\mathrm{TT}} \equiv \widetilde{C}_{k}^{\mathrm{\, TT}}+ N_{k}^{\mathrm{\, TT}}$ is the 
\emph{observed} temperature power APS as defined in Eq. (\ref{eq:obscltt}). 

Finally, the covariance of the integrated potential field estimator provides us with a four-point estimator 
of the integrated potential APS. When expanding the lensed CMB temperature modes at second
 order in $\phi$, the $\widehat \phi_{\mathrm{TT}}$ estimator covariance reads:
\begin{equation}
\langle \hat{\phi}_{\mathrm{TT}}(\vec L)\hat{\phi}_{\mathrm{TT}}(\vec L')\rangle \, = \,  (2\pi)^2\, \delta(\vec L+\vec L')\, C_{\mathrm{TT,TT}}^{\phi\phi}(L)\, ,
\label{eq:clddcov}
\end{equation}
where the estimated potential APS, $C_{\mathrm{TT,TT}}^{\phi\phi}(L)$, taking into account all 
sources of variances, both projected potential and CMB cosmic variance, instrumental noise and 
confusion noise from other potential Fourier modes, writes:
\begin{equation}
C_{\mathrm{TT,TT}}^{\phi\phi}(L) \, = \, C_{L}^{\phi\phi}+ N_{(0)}^{\, \phi\phi}(L) +
N_{(1)}^{\, \phi\phi}(L) + N_{(2)}^{\, \phi\phi}(L).
\label{eq:variance}
\end{equation}
Here, we have distinguished three different noise contributions to the
integrated potential estimator variance. The dominant noise
contribution, $N_{(0)}^{\, \phi\phi}(L) = A_{\mathrm{TT}}(L)$, depends only on the unlensed
 and the observed temperature APS. It represents the Gaussian contribution to the potential 
APS estimator, in a sense that it is the variance one would obtain by replacing the lensed
 temperature map in Eq.~(\ref{eq:estimator}), by a map with the same APS but Gaussian 
statistics. In addition, the potential APS estimator
suffers from sub-dominant non-Gaussian noise contributions. The first, quoted
$N_{(1)}^{\, \phi\phi}(L)$ has been calculated by~\citet{Kesden2003}. The second non Gaussian noise 
term $N_{(2)}^{\, \phi\phi}(L)$ is quadratic in $C_{L}^{\phi\phi}$. 
First calculated in~\citet{Hanson2009}, this term is shown to contribute even more than the 
first-order one at low multipoles ($L\leq 200$).

These two non Gaussian noise terms arise from the trispectrum part (or so-called connected part) of 
the four-point lensed temperature correlator hidden in the integrated potential field estimator covariance. 
It can be interpreted as the confusion noise coming from other integrated potential modes. Since it
depends on the integrated potential APS, which has to be
estimated, an iterative estimation scheme would be required for
taking it into account. However, our study based on simulated data
allows us to calculate these terms from the fiducial potential APS
 and then subtract it from the estimator variance.
 
%
%
%
%
%
%
%

\section{Effect of foreground removal: a Monte-Carlo analysis}
\label{sec:patches}

Up to now, no analysis has been performed to assess the effect of a
component separation process on the CMB lensing extraction. 
The question we propose to
 address here is whether the lensing signal is preserved in 
the CMB map output by the component separation process.
 In order to get a first insight, 
we use a Monte-Carlo approach within the flat sky approximation.
 

%
%
%
%
\subsection{Idealized {\sc Planck} sky model}
\label{sec:flatsimulation}

We create a simulation pipeline to generate some idealized synthetic patches of 
the sky for the {\sc Planck} experiment. Our sky model is a linear
uncorrelated mixture of \modif{the lensed CMB
temperature and astrophysical components, which includes} the
Sunyaev-Zel'dovich effect, the thermal emission of the interstellar
dust \modif{and the unresolved infrared point sources emission. In modeling
these three components, we ensure to catch the dominant foreground emission
features at the {\sc Planck}-HFI frequencies}. Then we add the nominal
effects of the {\sc Planck}-HFI instrument, modeled as a purely
Gaussian shaped beam and a spatially uniform white Gaussian noise.
Each hypothesis we adopt is a crude model of the astrophysical contaminant 
and systematic effects that pollute the {\sc Planck} data, and is intended to be a 
\emph{demonstration model} for a study devoted to the impact of the
component separation algorithms on the CMB lensing retrieval. 

We generate four sets of 300 {\sc Planck}-HFI synthetic patches of the
sky, with instrumental noise and, when needed, with foreground
emissions:

\begin{itemize}
\item{Set I} contains lensed CMB temperature maps generated from an
  unique fixed projected potential realisation and with the
  instrumental effects (beam and white noise);
\item{Set I-fg} is built from set I. In addition, a fixed realisation
  of dust and SZ is added to each set I map;
\item{Set II} is a set of lensed CMB temperature maps generated from
  300 random realisations of the lenses distribution plus the
  instrumental effects;
\item{Set II-fg} is built from set II. Each map of set II is
  superimposed with \modif{randomly picked} dust and SZ maps.
\end{itemize}

\modif{Note that the point sources emission will be included afterward in
our simulation pipeline, through a direct estimation of the point sources
residuals after component separation as described in
Sect.~\ref{sec:gmca}}. Sets I and I-fg will serve at studying the
projected potential \emph{field} reconstruction, whereas sets II and 
II-fg will be used in the projected potential \emph{angular power
  spectrum} (APS) estimates analysis. Hereafter, 
our method and its assumed hypothesis are detailed. 
 
%
%
%
\subsubsection{Lensed CMB temperature map}

Once we have assumed the Gaussianity of the integrated potential field, the lensed CMB temperature 
simulation principle is straightforward as a direct application of the remapping 
Eq.~(\ref{eq:remapping}). We start from the APS of both the temperature and the 
projected potential field as well as the cross-APS reflecting the correlation between the
 CMB temperature and the gravitational potential fields due to the Integrated Sachs-Wolfe (ISW) 
effect.
 Then we generate two Gaussian fields directly in the Fourier space, so that
%
\begin{eqnarray}
T(\vec{k})\, &=& \, \sqrt{C_{|\vec{k}|}^{\mathrm{TT}}} G_{(0,1)}^{\, (1)}(\vec k)\\\nonumber
\phi(\vec k) \, &=& \,  \sqrt{\frac{(C_{|\vec{k}|}^{\mathrm{T}\phi})^2}{C_{|\vec{k}|}^{\mathrm{TT}}}} G_{(0,1)}^{\, (1)}(\vec k)
+ \sqrt{C_{|\vec{k}|}^{\phi\phi} - \frac{(C_{|\vec{k}|}^{\mathrm{T}\phi})^2}{C_{|\vec{k}|}^{\mathrm{TT}}}} 
G_{(0,1)}^{\, (2)}(\vec k),
\label{eq:Tetphi}
\end{eqnarray} 
where $G_{(0,1)}^{\, (1)}(\vec k)$ and $G_{(0,1)}^{\, (2)}(\vec k)$ are two independent realisations of a
 Gaussian field of zero mean and unit variance. Because of the typical scales of the deflection 
field -- deflection angles are of the order of 2 or 3 arcmin (depending on the fiducial cosmological model)
 but correlated over several degrees 
on the sky -- the generated maps should be both high resolution and extended over not too small 
sky area. We choose to produce some $12.5\times12.5$ square degrees maps of 2.5~arcmin of 
resolution, as a good trade-off between the quality of the simulated maps and the time needed to
 the generation and the analysis of these maps.

\begin{figure}
  \resizebox{\hsize}{!}{\includegraphics[bb= 0 0 566 407,clip]{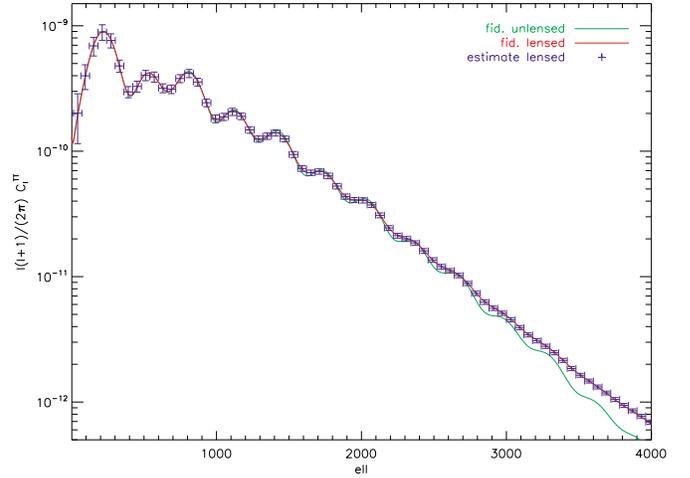}}
  \caption{CMB temperature APS. Red/black (respectively  green/grey) line is the lensed 
(respectively unlensed) temperature APS calculated with the public Boltzmann code 
{\sc camb}~\citep{Lewis1999,Challinor2005}. The blue/black data-points are the mean of the binned power 
spectrum reconstructed on 500 simulated lensed temperature maps. The error-bars are given by the 
variance of the 500 APS estimates.}
  \label{fig:lcltt}
\end{figure}

From CMB temperature and projected potential in the Fourier space, we calculate both the temperature
 and the deflection angles in the real space. The last step consists in performing the remapping of
 the primordial temperature map according to the deflection angles. Here is the technical point. 
Starting from a regular sample of a field (the underlying unlensed map), we have to extract an 
irregular sample of the same field (the lensed map) -- the new directions where to sample from are
 given by the previous one shifted by the deflection angles. Thus, this is a well-documented 
interpolation issue, the difficulty lying in the fact that the scale of the interpolation scheme 
is the same than the typical scale of the physical process under interest. We have to take 
particular care in the interpolation algorithm to avoid creating some spurious lensing signal or 
introducing additional non-Gaussianities. We find that a parametric cubic interpolation 
scheme~\citep{Park1983} apply reasonably well. In addition, to avoid any loss of power due to the 
interpolation, we overpixellize twice the underlying unlensed temperature and deflection field. 
The first test we perform to control the quality of the simulation is to compare the Monte-Carlo 
estimate of the APS over 500 simulations of the lensed maps with the analytical 
calculation of the lensed APS provided  by the 
{\sc camb}\footnote{web site: \tt{http://camb.info/}} Boltzmann code~\citep{Lewis1999, Challinor2005}.
 As shown in Fig.~\ref{fig:lcltt}, the APS of our simulated lensed maps is 
consistent with the theoretical one up to multipole 4000 -- which is largely enough to study the 
CMB lensing with {\sc Planck}.   
%
%
%
%

%
%
\subsubsection{Astrophysical components}
In any CMB experiment the temperature signal is mixed with foreground contributions
 of astrophysical origin -- among them we can separate the diffuse galactic emission (thermal and 
rotational dust, synchrotron, Bremsstrahlung (free-free) radiation) from the extragalactic 
components (point sources, thermal and kinetic Sunyaev-Zel'dovich effects). As discussed in the
 introduction, each of these components could potentially, if inefficiently removed, degrade our 
capability to reconstruct the CMB lensing. Here, to complete our demonstration sky model, we 
choose to simulate \modif{the dominant astrophysical foregrounds at
the {\sc Planck}-HFI frequencies, namely the thermal emission of the
galactic dust, the thermal SZ effect and the unresolved infra-red
point sources.} Thermal dust simulations are obtained from an interpolation of the
100$\mu$m IRAS data, in the sky region located around $\alpha = 204^{\circ}$ and 
$\delta = 11^{\circ}$, at the relevant CMB frequencies as described in~\citet{Delabrouille2003}. 
Note that several treatments have been applied on these maps -- point sources removal, destriping,
 inpainting in the Fourier space with constraint realisations -- which may 
induce an amount of additional non-Gaussianities. 
The SZ emission on the sky patches can be randomly selected in a set of 1500 
realisations produced with a semi-analytical simulation tool provided in the 
literature~\citep{Delabrouille2002}. Note that SZ emission is assumed
here not to correlate with the CMB lensed signal.
\modif{As for the estimates of the unresolved point sources residuals, we choose to take advantage 
of the refined full-sky simulations of the infra-red point sources emission in each {\sc Planck}-HFI 
frequencies, provided by the {\sc Planck} \emph{Component Separation Working Group} (WG2). In these 
simulations, the source counts are drawn from the IRAS catalog, and their spectral energy distributions 
are modeled following~\citet{Serjeant2005}. In addition, they involve several refinements,
such as the filling of the IRAS mask by synthetic data, the additional simulation of fainter sources 
according to the \citet{Granato2004} model, and their clustering (see~\citet{Leach2008}). Note that the 
radio-galaxies, another population of extragalactic sources, can be safely neglected here, as they lead 
to a sub-dominant emission compared to the infrared-galaxies one, at the {\sc Planck}-HFI frequencies.}

%
%
\subsubsection{{\sc Planck}-like noise}   

Finally, we simulate the effects of the {\sc Planck} High Frequency Instrument (HFI) according 
to their nominal characteristics~\citep{Bluebook2005}, which are summarized 
in Table~\ref{tab:planck}. At each frequency channel, the component mixture is convoluted by a 
Gaussian beam with the corresponding FWHM size. 
Then a spatially uniform white noise following a Gaussian statistic is added. Finally, the
 resulting maps can be deconvolved from the beam transfer function, resulting in an exponential 
increase of the noise at the scales corresponding to the beam size. Because smaller angular 
scale carry the larger amount of 
lensing information, the higher the angular resolution is, the better the lensing
 reconstruction can be. Our tests show that in the ideal case, the lensing reconstruction on 
{\sc Planck}-HFI synthetic maps is insensitive to the addition or the removal of the 100~GHz 
frequency channel information, whose beam function is roughly 
 twice larger than the beam in the higher frequency channels. Worst, when running the full 
Monte-Carlo chain, turning on the foreground emissions and the component separation process, 
adding the lower frequency channel results in increasing the confusion noise of the 
lensing reconstruction. That's why, we exclude the 100~GHz 
frequency channel of our analysis. To summarize, our {\sc Planck} sky model reads:
\begin{equation}
{\tens{T}^{\mathrm{obs}}} = \mathcal{A}\,{\tens{s}} \,+\, \mathcal{\tens{B}}^{-1} \otimes \tens{n}, 
\end{equation}
where $\tens{T}^{\mathrm{obs}}$ is the set of five individual frequency channel maps, $\mathcal{A}$ is the 
mixing matrix calculated from the frequency dependence of the three signal contributions, $\tens{s}$ is the 
set of CMB, dust and SZ maps at a pivot frequency, ${\mathcal{\tens{B}}^{-1}}$ is the set of Gaussian beam 
inverse transfer functions and {\tens{n}}, the set of white noise maps in the five frequency channels we have 
selected. Note that we assume a perfect beam deconvolution process.
  
\begin{table}
  \caption{Instrumental characteristics of {\sc Planck}-HFI$^{\mathrm{a}}$}
   \label{tab:planck}
   \centering
    \begin{tabular}{ccc}
        \hline\hline
        \noalign{\smallskip}
        channel (GHz) & $\theta_{\mathrm{fwhm}}$ (arcmin) & $\sigma_{\mathrm{T}}$ ($\mu$K$^2$.sr$^{\, -1}$) \\
        \noalign{\smallskip}
        \hline
        \noalign{\smallskip}
         100 &  9.5 & 6.8    \\
         143 &  7.1 & 6.0    \\
         217 &  5.  & 13.1   \\
         353 &  5.  & 40.1   \\
         545 &  5.  & 401    \\
         857 &  5.  & 18\,300. \\ 
        \noalign{\smallskip}
        \hline
    \end{tabular}

\begin{list}{}{}
\item[$^{\mathrm{a}}$] \citep[see][]{Bluebook2005}
\end{list}
\end{table}

%
%
%
%
%
\subsection{Component separation using GMCA}
\label{sec:gmca}

\indent For most of the cosmological analysis of the CMB data -- and for the CMB lensing extraction in particular -- 
the cosmological signal has to be carefully disentangled from the other sources of emission that 
contribute to the observed temperature map. The component separation is a part of the signal 
processing dedicated to discriminate between the different contributions of the final maps. Briefly, 
the gist of any component separation technique consists in taking profit of the difference in the 
frequency behavior and the spatial structures (i.e. morphology) that distinguish these different 
\emph{components}.
 From a set of frequency channel maps, a typical component 
separation algorithm provides a unique map of the CMB temperature with the instrumental noise and a 
foreground emission residual. In general, the lower the foreground residual rms level the better the 
separation algorithm. However, this simple rule is not necessarily true for CMB lensing 
reconstruction. In this case, preserving the statistical properties of the underlying CMB temperature 
map is critical. 

\indent In the {\sc Planck} consortium, the component separation is a critical issue, involving a 
whole Working Group (WG2) devoted to provide several algorithms for separating 
CMB from foregrounds and to compare their merits (see \citet{Leach2008} for a recent comparison of 
the current proposed methods). Eight teams have provided a complete component separation pipeline 
capable to treat a realistic set of {\sc Planck} temperature and polarisation maps. Each methods 
differ in the external constraints they use, the physical modeling they assume and the algorithm 
they are based on.

%
\begin{figure*}[t]
  \centering
  \vfill
  \includegraphics*[width=155pt,height=145pt]{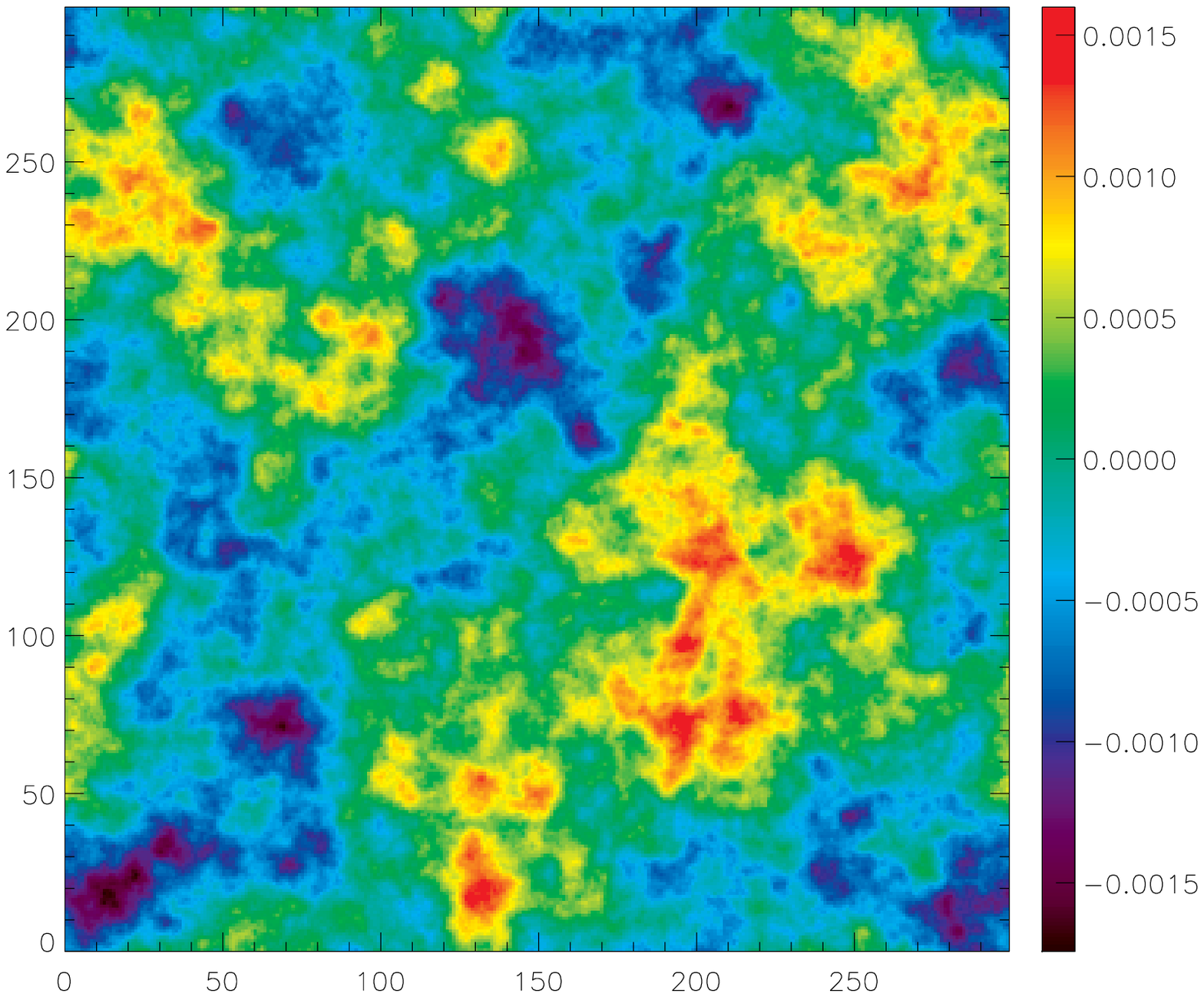}
  \hfill
  \includegraphics*[width=145pt,height=145pt]{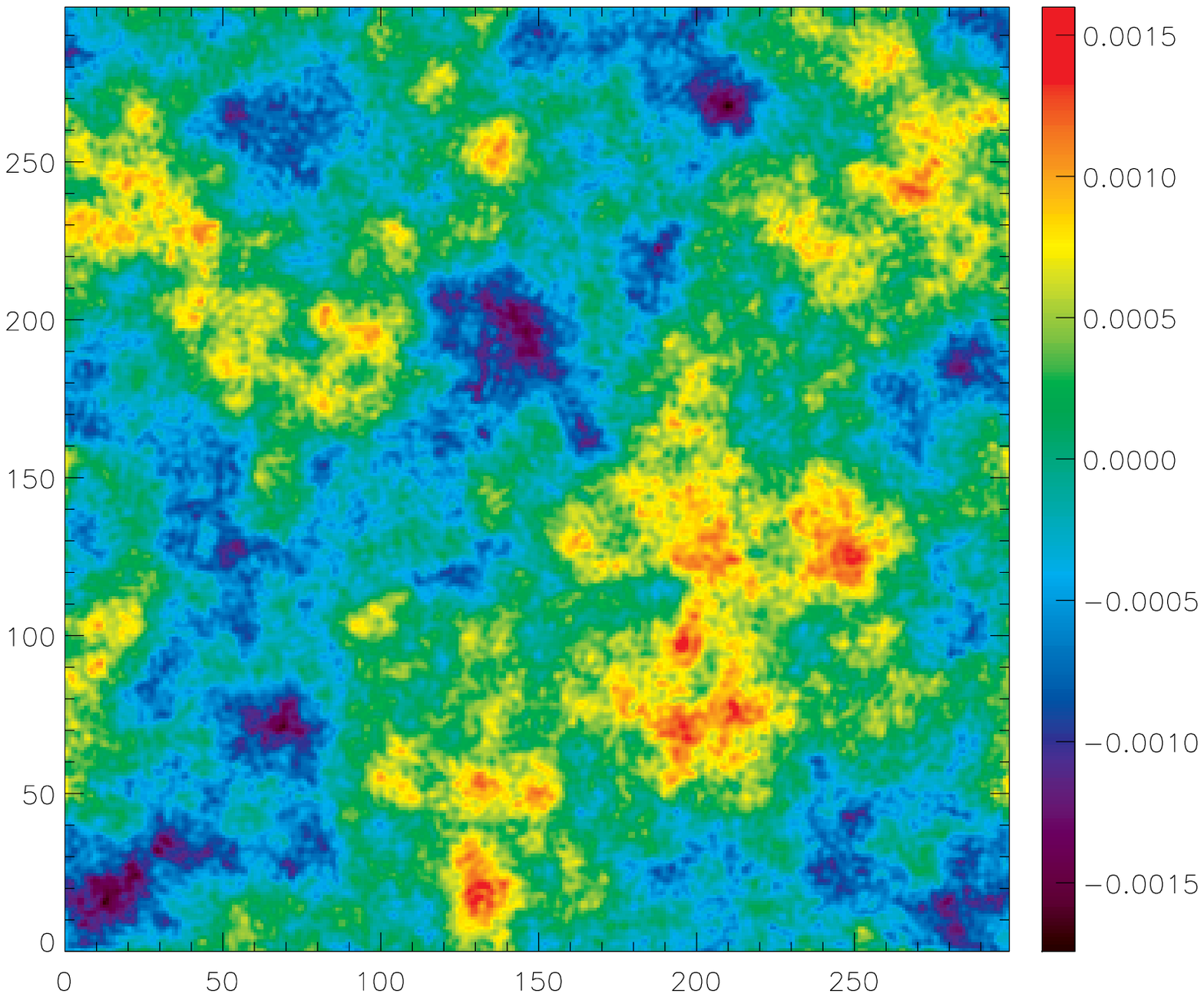}
  \hfill
  \includegraphics*[width=170pt,height=145pt]{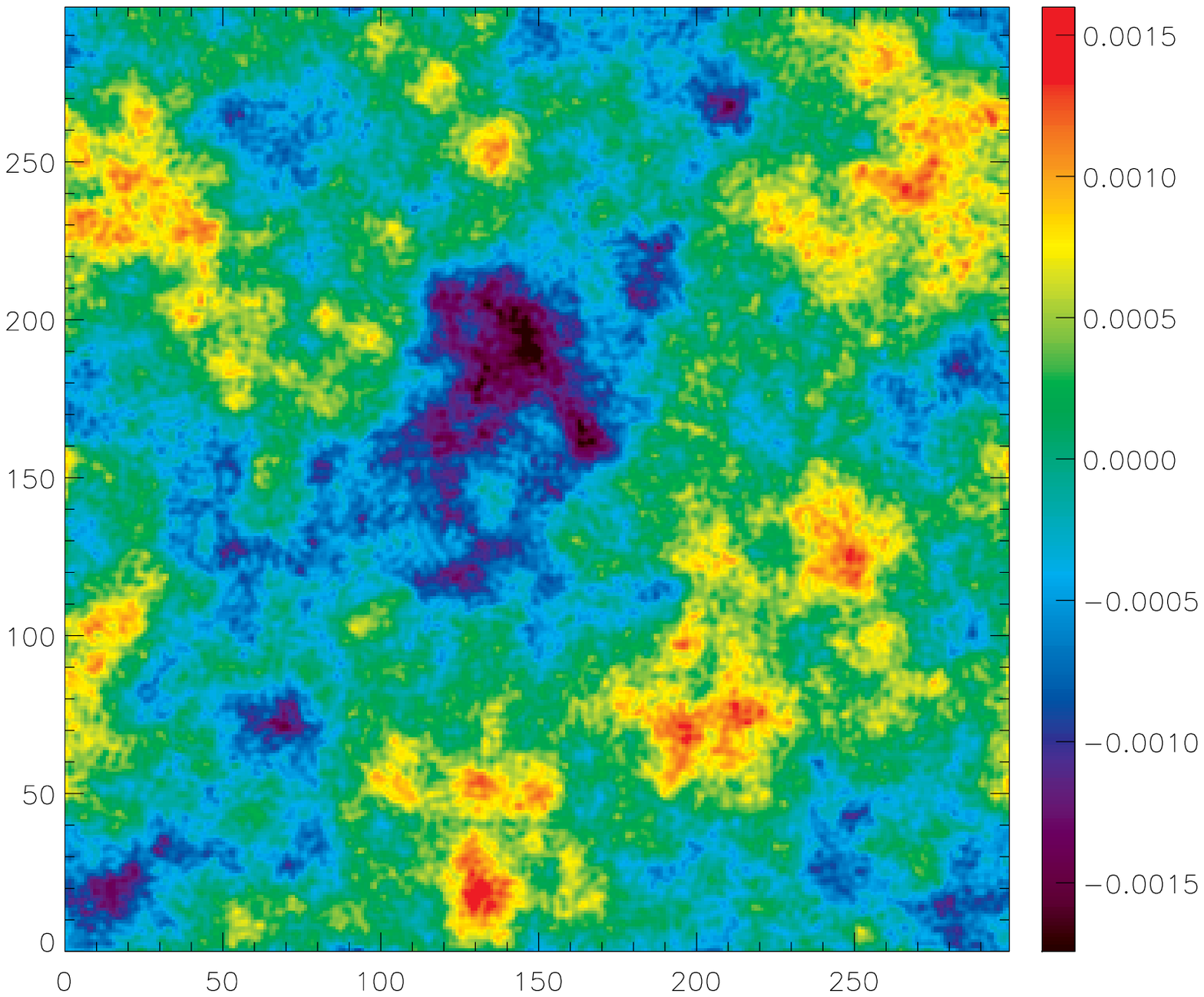}
  \vfill
  \caption{Impact of the foregrounds residuals on the deflection field reconstruction 
on $12.5^\circ \times 12.5^\circ$ square patches. Left: the input realisation of the deflection amplitude; 
Middle: the stack of 300 deflection estimates from the set~I maps 
(synthetic {\sc Planck} temperature maps -- without any foregrounds residuals); Right: the stack 
of 300 deflection estimates from the set~I-{\sc gmca} maps ({\sc Planck} temperature 
maps output of the GMCA component separation process). All maps have the same color table shown in 
the right-most part of the figure.} 
  \label{fig:phimaps}   
\end{figure*}
%

\indent Among the available techniques, we choose to use the {\emph Generalized Morphological 
Component Analysis} (hereafter GMCA) which is a blind method component separation method. In GMCA, 
each observation 
$\mathrm{T}^{\mathrm{obs}}_\nu(\hat{\vec n})$ is assumed to be the linear combination of $n_\mathrm{c}$ 
\emph{components} $\{\mathrm{s}^i(\hat{\vec n})\}_{i=1,\cdots,n_\mathrm{c}}$ such that:
\begin{equation}
\mathrm{T}_\nu^{\mathrm{obs}}(\hat{\vec n}) \, = \, \mathcal{A}_{\nu i}\, \mathrm{s}^{i}(\hat{\vec n}) \,+\, \mathrm{n}_\nu(\hat{\vec n}),
\end{equation}
 where $\mathrm{n}_\nu(\hat{\vec n})$ models instrumental noise. The general idea subtending this algorithm 
is the fact that the components, which result from completely different physical processes, have 
different spatial morphologies or structures. These morphological differences translate into a 
difference in their representation into a fixed waveform dictionary $\mathcal{D}$. If only a few 
coefficients of a fixed dictionary are enough to completely represent a given component, this 
component is said to be sparse in that dictionary $\mathcal{D}$. The dictionary succeeds in catching
 the general features that characterize the component morphology. That's why, separating the observed 
map into components that maximize their sparsity in a given dictionary is an efficient strategy to 
distinguish between physically different emission sources. In practice, a wavelet basis is a good 
choice for astrophysical components that overwhelmingly contain smooth spatial features. GMCA is a 
sparsity-maximization algorithm, a notion that we briefly introduce hereafter. 
Let $\{{\rm{d}}^j(\hat{\vec n})\}$ be the set of vector that forms the dictionary $\mathcal{D}$. 
Let $\alpha_{ij} = \langle {\rm{s}}^i(\hat{\vec n}) , {\rm{d}}^j(\hat{\vec n}) \rangle$ denote the 
scalar product coefficients between ${\rm{s}}^i(\hat{\vec n})$ and ${\rm{d}}^j(\hat{\vec n})$. When 
$\mathcal{D}$ is an orthogonal wavelet basis, the following properties hold~:
\begin{eqnarray}\nonumber
\langle {\rm{d}}^i(\hat{\vec n}) , {\rm{d}}^j(\hat{\vec n}) \rangle & = & 0\mbox{ if } i \neq j \\\nonumber
\langle {\rm{d}}^i(\hat{\vec n}) , {\rm{d}}^i(\hat{\vec n}) \rangle & = & 1     \\\nonumber
 {\rm{s}}^i(\hat{\vec n})  & =  & \sum_j \langle {\rm{s}}^i(\hat{\vec n}) , {\rm{d}}^j(\hat{\vec n}) \rangle \, {\rm{d}}^j(\hat{\vec n}) 
\end{eqnarray}
 Then GMCA estimates the components $\{ \mathrm{s}^i(\hat{\vec n})\}$ and the mixtures weights 
$\{\mathcal{A}_{\nu i}\}$ by maximizing the sparsity of each component in $\mathcal{D}$. As 
advocated in \citet{Bobin2008}, a good sparsity estimate is the sum of the absolute values of 
$\{\alpha_{ij}\}_{i,j}$. Maximizing the sparsity of the components is then equivalent 
to minimizing this sparsity measure. The model parameters are estimated by GMCA as follows:
\begin{equation}
\min_{\{ \mathrm{s}^i(\hat{\vec n}) \},\{ \mathcal{A}_{\nu i}\}} \sum_{ij} \left |\alpha_{ij}\right| \mbox{s.t.} \left 
\| \mathrm{T}_\nu^\mathrm{obs}(\hat{\vec n}) -  \mathcal{A}_{\nu i}\mathrm{s}^i(\hat{\vec n}) \right \| < \epsilon 
\end{equation}
where $\epsilon$ stands for the reconstruction error.  The norm $\| \, . \, \|$ stands for the 
usual $\ell_2$ norm:
\begin{equation}
\left\| \mathrm{T}_\nu^{\mathrm{obs}}(\hat{\vec n}) -  \mathcal{A}_{\nu i}\mathrm{s}^i(\hat{\vec n}) \right \| = 
\sqrt{\sum_{\nu,i,\hat{\vec n}} (\mathrm{T}_\nu^\mathrm{obs}(\hat{\vec n}) -  \mathcal{A}_{\nu i}\mathrm{s}^i(\hat{\vec n}) )^2 }
\end{equation}
GMCA estimates the components $\{\mathrm{s}^i(\hat{\vec n})\}$ which have only a few significant 
coefficients $\{\alpha_{ij}\}$ in the dictionary; i.e. the components which are sparse in 
$\mathcal{D}$. Further technical details are given in~\citet{Bobin2008}.\\
For {\sc Planck}, the parameter $\epsilon$ is chosen to be very small. In that case, 
the components $\{ \mathrm{s}^i(\hat{\vec n})\}$ are estimated by applying the pseudo-inverse of the
 mixing matrix $\mathcal{A}$ to the observation channels 
$\{ \mathrm{T}_\nu^{\mathrm{obs}}(\hat{\vec n}) \}$:
\begin{equation}
\label{eq:compestim}
\mathrm{s}^i(\hat{\vec n}) = \sum_{\nu} \mathcal{A}^{+}_{\nu i} \mathrm{T}_\nu^\mathrm{obs}(\hat{\vec n}) 
\end{equation}
where $\mathcal{A}^+_{\nu i}$ is the element at position $\{i,\nu \}$ of the pseudo-inverse matrix 
of $\mathcal{A}$ defined as: 
$\mathcal{A}^+ = \left( \mathcal{A}^\mathrm{T} \mathcal{A}\right)^{-1}\mathcal{A}^\mathrm{T}$. Interestingly, 
the contribution of the component separation is then linear. As a consequence, the noise perturbing
 each component can be accurately known. Furthermore, the linearity of the separation guarantees
 that the separation technique itself does not generate non-Gaussianity in estimated CMB map. 
Only the residual terms coming from the other components can create non-Gaussian features in the CMB.

\modif{Another important consequence is that these properties give us a conservative method to estimate the 
point sources residuals remaining within the CMB maps after foreground cleaning, as describe hereafter. 
Because each source has its own spectral property, component separation techniques fail at disentangling 
the point sources emission from the observed maps. As a result, point sources remain mixed with the other 
components and the precise amount of the point sources emission by observation channels which has leaked 
in each component, is determined by the coefficients of the mixing matrix. More formally, in order to estimate
 the point sources residuals embedded in the foreground-cleaned CMB maps, quoted 
$\mathrm{s}^\mathrm{ps}(\hat{\vec n})$, one can apply Eq.~\ref{eq:compestim} to the simulated point sources 
in the observation channels $\{ \mathrm{T}_\nu^{\mathrm{ps}}(\hat{\vec n}) \}$:  
\begin{equation}
\mathrm{s}^\mathrm{ps}(\hat{\vec n}) = \sum_{\nu} \mathcal{A}^{+}_{\nu 0} \mathrm{T}_\nu^\mathrm{ps}(\hat{\vec n}) 
\end{equation}
where the elements $\{ \mathcal{A}^+_{\nu 0} \}$ form the column of the pseudo-inverse matrix which corresponds
 to the CMB component. Then the brighter point sources, which have been previously detected in the 
{\sc Planck}-HFI channels, are masked out and the corresponding gaps are restored using an 
\emph{inpainting} method. A detailed description of the mask and the restoration technique will be given 
hereafter in the Sect.~\ref{sec:fullsky}. The final full-sky map we obtain is an estimate of the unresolved 
infrared point sources residuals, which contaminate the CMB temperature map after component separation with 
GMCA. This sky map is divided into patches of $12.5 \times 12.5$ square degrees with $50\%$ overlapping. 
We form a set of 300 point sources residuals square maps by selecting the patches with a maximal $30\%$ masked 
area.}

We perform the component separation using GMCA on sets I-fg and II-fg, each of 300 simulated patches 
generated following our idealized {\sc Planck} sky model and described in the previous section 
(\ref{sec:flatsimulation}). As an output of this process, we obtain two sets of 300 foreground-cleaned 
CMB temperature maps. \modif{Note that GMCA achieves the extraction of the foreground components as well. 
Unresolved point sources residuals are added to each map of these two sets. In the following, we will refer 
to the sets of lensed CMB maps with galactic dust, SZ effect and point sources residuals after the GMCA component 
separation, as sets I-{\sc gmca} and II-{\sc gmca} respectively.}

%
%
%
%
%
\subsection{CMB lensing reconstruction} 
\label{sec:flatestimation}

Here, we apply a discrete version of the quadratic estimator written by Okamoto $\&$ Hu 
(Eq.~\ref{eq:estimator}) on the different sets of simulated maps previously described, namely 
sets I, I-{\sc gmca}, II and II-{\sc gmca}. We seek to assess the foreground residuals impact 
on our capability to reconstruct CMB lensing with {\sc Planck}. 

%
%

\subsubsection{Testing the estimator performances}
\label{sec:idealrec}

First, we give explicitly the expression of the discrete quadratic estimator that 
we derive from Eq.~(\ref{eq:estimator}): 
\begin{equation}
\hat{\phi}_{\mathrm{TT}}(\vec U) =  \frac{A_{\mathrm{TT}}(L)}{L^2 \tens{A}} 
\sum_{\mathcal{D}_1 \cap \mathcal{D}_2} \bar{T}(\vec u_1) \bar{T}(\vec u_2)  
F_{\mathrm{TT}}(\vec u_1,\vec u_2)\,,
\label{eq:discresti}
\end{equation}
where $\vec U$, $\vec{u_1}$ and $\vec u_2$ are wave-vectors related by $\vec{u_1}+\vec{u_2} = \vec U$ and
 $\tens{A}$ is the area of the sky patch considered. The sum is performed on the intersection of 
two disks $\mathcal{D}_1$ and $\mathcal{D}_2$. The former is the zero-frequency centered disk defined
 by $\Delta_1 |\vec{u_1}| \leq L_{\mathrm{max}}$, where ${\Delta_1}$ denotes the frequency interval in the 
Fourier space ( i.e. the smallest nonzero positive frequency), which equals $2\pi/\sqrt{\tens{A}}$. 
The latter, namely $\mathcal{D}_2$, is also a disk 
of radius $L_{\mathrm{max}}$ but centered around $\vec U$. It is therefore defined by 
$\Delta_1|(\vec u_2 - \vec U)| \leq L_{\mathrm{max}}$. The wave-vectors $\vec u_1$, $\vec u_2$
and $\vec U$ correspond to $\vec k_1 / \Delta_1$, $\vec k_2 / \Delta_1$ and 
$\vec L / \Delta_1$ respectively. The normalisation $A_{\mathrm{TT}}(L)$ and the weighting function 
$F_{\mathrm{TT}}(\vec u_1,\vec u_2)$ are respectively the discrete version of Eq.~(\ref{eq:att}) 
and Eq.~(\ref{eq:optimal}). For the {\sc Planck}-HFI experiment, we verify that the reconstructed potential 
field does not vary either we cut the sum in Eq.~(\ref{eq:discresti}) at $L_{\mathrm{max}} = 2600$ or we push
 it further. 
 
We study our capability to reconstruct a map of the integrated potential field with 
{\sc Planck}-HFI idealized simulation, assuming a perfect component separation without any foreground
 residuals. We apply the discrete quadratic estimator on set I maps (see Sect.~\ref{sec:flatsimulation})
 to obtain 300 estimates of the same realisation of the projected potential field $\phi$. 
Once stacking these estimates, the final $\phi$ map is an estimate of the input $\phi$ realisation.
\modif{Following~\citet{Hu2001a}, we prefer to present our results in terms of the deflection field amplitude
 rather than the very smooth gravitational potential field, in order to highlight the intermediate angular 
scales features. Fig.~\ref{fig:phimaps} shows the input deflection field realisation, which has been used to 
simulate the lensing effect in the set I maps (on the first panel), as well as its reconstruction with the 
quadratic estimator applied on the set I maps (second panel). Even if the reconstruction noise is visible at 
smaller angular scales, the features of the deflection map are well recovered.} 
   
Characterizing {\sc Planck} sensitivity to the projected potential APS requires 
to account for both the CMB and the projected potential field cosmic variances. Thus, we move on to set II.
 As previously, we apply the quadratic estimator (Eq.~\ref{eq:discresti}) on the lensed CMB maps to reconstruct 
projected potential fields. Averaging over the APS of these individual $\phi$ field estimates gives
 an evaluation of the quadratic estimator variance (as defined in Eq.~\ref{eq:clddcov}). The final reconstructed 
projected potential APS is obtained in subtracting the noise contributions, 
described in Sect.~\ref{sec:formalism}, from the variance. The former is related by Eq.~(\ref{eq:dd2phiphi}) to 
the deflection APS shown in the Fig.~\ref{fig:cldd}. The error bars are estimated as the dispersion
 between each individual deflection APS reconstruction. Thus, set II maps, which are idealized versions 
of the {\sc Planck}-HFI sky assuming a perfect component separation, lead to a good reconstruction of the 
deflection APS up to $L_{\mathrm{max}}=2600$. The error-bars evaluated here give an upper limit of the 
{\sc Planck}-HFI sensitivity to the deflection APS. As one can see on Fig.~\ref{fig:ngnldd}, they are compatible 
with the theoretical 1$\sigma$ error-bars one can calculate from the Fisher formalism:
\begin{equation}
\Delta C_L^{\mathrm{dd}} = \sqrt{\frac{1}{N_\mathrm{eff}}}(C_L^\mathrm{dd}+N_L^\mathrm{dd}),
\label{eq:fisher_error}
\end{equation}
where $N_\mathrm{eff} = 4\pi / L \Delta L \tens{A}$ is a naive estimate of the independent available
 Fourier modes. The error bars estimated here will provide us with a comparison level to quantify the impact of 
the foreground residuals. 

%
%

\subsubsection{Impact of the foreground residuals}

Here we essentially redo the same analysis but using the full-simulation pipeline of our {\sc Planck}-HFI 
demonstration model. The integrated potential field is 
extracted from the sets I-{\sc GMCA} and II-{\sc GMCA} described in Sect.~\ref{sec:gmca}.

First, we aim at developing an intuition on the impact of foreground residuals on the deflection map 
reconstruction. We use the set I-{\sc gmca}, in which both deflection field and foregrounds realisations are 
fixed. As previously, the 300 deflection field estimates reconstructed with the quadratic estimator are stacked
 to produce an unique reconstructed deflection field shown in Fig.~\ref{fig:phimaps}. We can see that the 
recovery of the underlying deflection field is still achieved even in presence of foreground emission and after
 the GMCA analysis. \modif{The impact of the foreground residuals is nevertheless visible, mostly at angular 
scales larger than 2 degrees whereas the intermediate angular scale features seem more preserved.}

For a more quantitative analysis, we move on to the impact of the foreground residuals on the deflection APS
 reconstruction. We use the set II-{\sc gmca} (see Sects.~\ref{sec:flatsimulation} and \ref{sec:gmca}) 
to ensure that the variances of the CMB, the deflection field 
and the foregrounds are accounted for. We reconstruct a deflection field estimate from each of the 
set~II-{\sc gmca} maps using the quadratic estimator given in Eq.~(\ref{eq:discresti}). Finally, we obtain the 
reconstructed deflection APS from the average variance over these 300 deflection APS estimates as described in 
Sect.~\ref{sec:idealrec}.  
%
\begin{figure}
  \resizebox{\hsize}{!}{\includegraphics[bb= 7 0 555 406,clip]{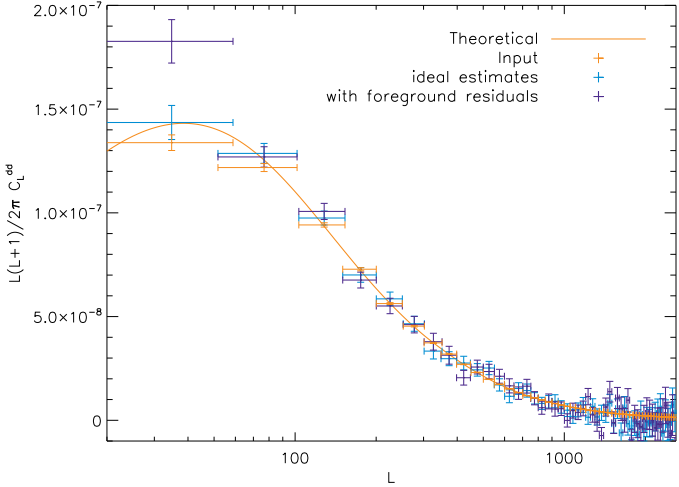}}
  \caption{The deflection APS. Data-points are the binned APS reconstructed from {\sc Planck} 
synthetic lensed CMB maps in two cases: (light blue/grey) the \emph{ideal} case without any foregrounds and 
(dark blue/black) the case with foregrounds residuals from the GMCA output CMB maps. The fiducial deflection APS
 calculated with {\sc camb} is figured by the (orange/solid) line; Orange/grey data points are the binned deflection APS
 estimates on the 300 input deflection field realisations. The horizontal and vertical intervals associated 
with the data points represent the averaging multipole bands and the 1$\sigma$ errors respectively.}
  \label{fig:cldd}
\end{figure}

\begin{figure}
  \resizebox{\hsize}{!}{\includegraphics[bb= 7 0 555 406,clip]{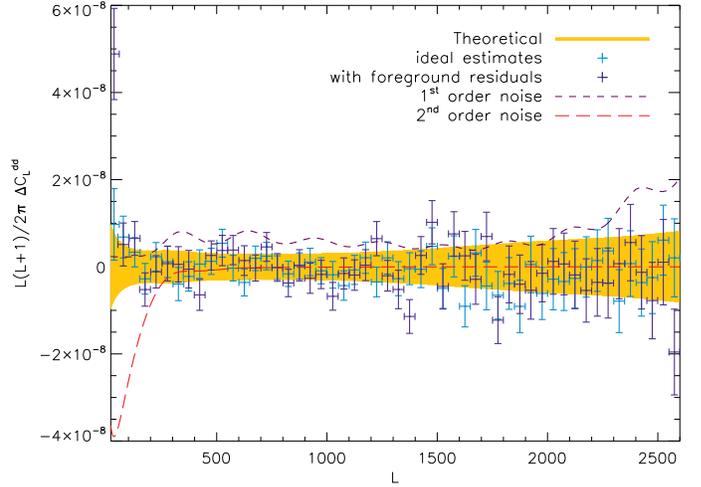}}
  \caption{Residual bias of the deflection APS reconstruction. Data-points figure the 
difference between the reconstructed and the input deflection APS averaging over 300 estimates: 
in the \emph{ideal} case in light blue/grey and in the case with foreground residuals in dark blue/black (consistently 
with the Fig.~\ref{fig:cldd} caption). The lines show the non-Gaussian noise terms of the 
quadratic estimator, at the first-order (violet/dashed) and at the second-order (red/long dashed) in $C_L^{\mathrm{dd}}$. 
Orange/grey colored band is the analytical 1$\sigma$ error band derived from the Fisher formalism.}  
  \label{fig:ngnldd}
\end{figure}

The reconstructed binned deflection APS with the evaluated 1$\sigma$ errors is 
represented in the Fig.~\ref{fig:cldd}. Fig.~\ref{fig:ngnldd} shows the difference between the 
reconstructed and the input deflection APS. First we report that the foreground residuals does 
not \modif{compromise} the {\sc Planck}-HFI capability to reconstruct the deflection APS -- or equivalently 
the integrated potential APS. Fig.~\ref{fig:cldd} shows that the APS reconstruction is preserved 
\modif{at the angular scales from $L=60$ up to $L=2600$.} In this multipole range, the GMCA algorithm succeeds 
in letting unchanged the statistical properties of the lensed CMB temperature anisotropies, which suggests 
that this is a well-appropriated component separation tool for CMB lensing reconstruction. 
\modif{As for the first multipole bin, we report a $4\sigma$ excess of the deflection signal in the 
$L=2$ to $L=60$ multipole range. We have checked that this bias is linked to the introduction of 
the unresolved point sources residuals in our simulation pipeline. Interestingly, we find that this excess 
originates not from the level of residuals themselves, but mostly from the cutting procedure}\footnote{%
The cutting procedure involves a sphere-to-plan projection, an apodisation and a fit of the Fourier coefficients 
of the square map. From the tests we run (not presented here), the apodisation appears to have the most 
harmful impact on the high angular scale deflection reconstruction. A complete study of the impact of the 
sphere to patches transition will be the subject of a companion paper.} %
\modif{we use to extract the set of 300 square maps from our full-sky point sources residuals. We postpone a 
closer inspection of the low multipole lensing reconstruction behavior to the complete full-sky study 
(hereafter in Sect.~\ref{sec:fullsky}). Apart 
from the excess signal in the first bin, the impact of the foreground residuals on the deflection reconstruction
 is also slightly visible at all angular scales, as 
seen in the Fig.~\ref{fig:ngnldd}.} If the difference between reconstructed and input APS is still compatible 
with zero within the theoretical 1$\sigma$ errors in the 60 to 2600 multipole range, this residual bias appears
 more featured, more oscillating than in the previous no foreground case.
To quantify this degradation in the deflection APS reconstruction, we calculate the total error in unit of
 $\sigma$, defined as:
\begin{equation}
 \Delta = \sum_{b} | \frac{\hat{C}_b^{\mathrm{\, dd}} - C_b^{\mathrm{\, dd}}}{\sigma_b}|,
\end{equation}
where $\hat{C}_b^{\mathrm{\, dd}}$ and $C_b^{\mathrm{\, dd}}$ are respectively the reconstructed and input 
deflection APS in the $b$ frequency band, and $\sigma_b$ the 1$\sigma$ error on $\hat{C}_b^{\mathrm{\, dd}}$. 
\modif{Note that when evaluating the total error, we exclude the first bin bias, which has been discussed 
previously.} With this definition, we find a total error, \modif{$\Delta_{\mathrm{ideal}} = 33$}, in the no 
foregrounds case whereas \modif{$\Delta_{\mathrm{gmca}} = 41$}, in presence of GMCA foreground residuals. 
\modif{If the total error serves as quantifying the increase of the \emph{bias} of the deflection APS 
reconstruction, one can also 
evaluate the increase of the \emph{errorbars}. We find that the presence of foreground residuals results 
in a $10\%$ increase of the errorbars on average}. It suggests that at least an amount of the non-Gaussian
 foreground residuals is mixed up with the lensing signal in the reconstruction process. Foregrounds showing 
some small angular scale (around 15 to 5 arcmin) features -- such as concentrated dust emission or brightest
 SZ clusters and point sources smeared by the instrument beam function -- are potentially the more challenging
 for the CMB lensing reconstruction. \modif{In addition, we remind that our sky model is intended to catch the 
dominant foreground features at the {\sc Planck}-HFI frequencies. The sub-dominant existing emissions, as free-free
 or synchrotron emissions, may marginally degrade our results as they are expected to slightly increase the foregrounds
 residuals whithin the CMB map.} 
 
\modif{As a final remark, we note that, because of the first bin excess signal problem, linked to the sphere 
to patches transition, one might privilege a full-sky approach when seeking at a precise lensing reconstruction
 at the higher ($L<60$) angular scales.}

%
%
%
%
%
%
%

\section{Impact of masks on the full-sky lensing reconstruction}
\label{sec:fullsky}

\subsection{Introduction}

From now, we move to a full-sky analysis of the CMB lensing effect. 
Some large area of the map, where the CMB signal is highly dominated by the foreground emission 
(e.~g. the galactic plan, the point sources directions), have to be masked 
out. Cutting to zero introduces some mode-coupling within the CMB observables. 
As the lensing reconstruction methods rely on the off-diagonal terms of the CMB data covariance matrix, 
the map-masking yields some artifacts in the projected potential estimate if not accounted for. Several
 methods have been proposed to treat the masking effect for extracting the lensing potential
 field from the WMAP data. In \citet{Smith2007}, the reconstruction is performed on the least-square estimate 
of the signal given the all-channels WMAP temperature data, requiring the inversion of the total data 
covariance matrix $(S+N)$. However, as relying at least on the inversion of the noise covariance 
matrix, such an optimal data filtering approach is very CPU-consuming when applied to the WMAP
maps. The {\sc Planck}-HFI resolution enables to provide 50 Mega pixels maps. Thus, 
the previous method to account for the masking will be difficult to extend to {\sc Planck}. In 
\citet{Hirata2008}, the need for dealing with the noise covariance matrix is avoided by 
cross-correlating different frequency band maps. However, this method implies that no component
 separation has been performed on the CMB maps before the lensing extraction. As a result, a 
lot of non-Gaussianities of foreground emission origin yields some artifact in the projected potential 
estimate, requiring a challenging post-processing to be corrected out. As previously mentioned, 
the {\sc Planck} collaboration devotes strong efforts in 
the component separation activities and the current developed methods have already proved their 
efficiency~\citep{Leach2008}. Moreover, the results we obtained with the 
\emph{demonstration analysis} (see Sect.~\ref{sec:patches}) tend to indicate that the lensing reconstruction is 
still doable after a component separation. Hence we plan to exploit the {\sc Planck} frequency band 
maps to clean out the foreground emission before reconstructing the lensing potential rather than using the 
cross-correlation based lensing estimator. As a consequence, we need an alternative 
method to solve the masking issue in maps at the {\sc Planck} resolution. Here, we propose to use an 
\emph{inpainting} method, assess its impact on the CMB lensing retrieval \modif{and check its robustness to 
the presence of foregrounds residuals within the CMB map.}
 
First, we describe the hypothesis assumed and the tools we use to generate synthetic all-sky lensed 
temperature maps for {\sc Planck}.
Then we describe our full-sky lensing estimator and test its performances on some {\sc Planck}-like 
temperature maps. Finally, we review the \emph{inpainting} method and conclude in studying the effect of the 
\emph{inpainting} on the projected potential APS reconstruction \modif{in two cases, first assuming a perfect
 component separation then in presence of point sources residuals.}

\subsection{Full-sky simulation}
\label{sec:spheresimu}

The formalism reviewed in Sect.~\ref{sec:formalism} is almost fully applicable
 to the spherical case. In particular, the remapping 
equation (Eq.~\ref{eq:remapping}) still holds, so that a lensed CMB sphere is given by:
\begin{equation}
  \widetilde{T}(\hat{\vec n}) = T(\hat{\vec n} + {\mathbf \nabla} \phi(\hat{\vec n})),
  \label{eq:remapsphere}
\end{equation} 
where the ${\bf \nabla}$ operator is to be understood as the covariant derivative on the 
sphere~\citep{Lewis2006}. ${\bf \nabla} \phi$ 
is identified to be the deflection field $\vec d$. Its zenithal and azimuthal coordinates 
can be calculated as the real and imaginary parts of a complex spin one field, using
spin-weighted spherical harmonics transform -- the detailed calculation can be found 
in~\citet{Hu2000} and \citet{Lewis2005}.

The {\sc LensPix}\footnote{{\tt http://cosmologist.info/lenspix/}} package 
described in \citet{Lewis2005} aims at generating a set of lensed CMB
temperature and polarisation maps from the analytical auto- and cross-APS of 
$\{T,E,B,\phi\}$, the temperature, the E and B polarisation modes and the line-of-sight projected 
potential respectively. The maps are provided in the 
{\sc HEALPix}\footnote{The acronym for 'Hierarchical Equal Area isoLatitude Pixelization' of 
a sphere (see {\tt http://healpix.jpl.nasa.gov/index.shtml}).} 
pixelisation scheme~\citep{Gorski2005}. The CMB lensing simulation is achieved in remapping the
 anisotropy fields according to Eq.~(\ref{eq:remapsphere}) of a higher
 resolution map using a bi-cubic interpolation scheme in equi-cylindrical pixels. A lensed temperature
map at the {\sc Planck} resolution (nside = 2048), can be computed  in
about 5~minutes on a  4-processors machine. 
The relative difference between the lensed temperature APS 
reconstructed on such a map and the analytical lensed APS obtained with 
{\sc CAMB}\footnote{The 'Code for Anisotropies in the Microwave Background' is a so-called
 Boltzmann's code described at {\tt http://camb.info/}}~\citep{Lewis1999} is below 1$\%$ up 
to $l=2750$. Here, to conservatively ensure a relative error below 1$\%$, we choose a multipole 
cut at $L_{\rm{max}} = 2600$.  By its speed and precision quality, {\sc LensPix} is a well adapted tool for 
a CMB lensing analysis with {\sc Planck} data alone.      

We obtained some {\sc Planck}-HFI synthetic maps as the flat-sky case (see Sect.~\ref{sec:patches}). 
White Gaussian noise realisations and Gaussian beam effect are 
added to the lensed temperature maps provided by the {\sc LensPix} code. This 
Gaussian noise contribution is fully defined by the all-channels beam-deconvolved APS given by:
\begin{equation}
N_l^{\mathrm{TT}} = \left( \sum_{\nu}\frac{1}{N_l^{\mathrm{\, TT},\nu}}\right)^{-1},
\label{eq:spherenltt}
\end{equation}
where $N_l^{\mathrm{\, TT},\nu}=(\theta_{\mathrm{fwhm}} \, \sigma_T)^2 \,\, exp\left[ l(l+1) \theta_{\mathrm{fwhm}}^2/8\ln{2}\right]$, 
$\theta_{\mathrm{fwhm}}$ and $\sigma_\mathrm{T}$ are the full width at half maximum of the beam and the level of
 white noise per resolution element respectively, as given in Table~\ref{tab:planck}.
The noise map is generated using the map creation tool of the {\sc HEALPix} package.

For  the lensing reconstruction analysis, we prepare 
two sets of 50 all-sky maps with 1.7~arcmin of angular resolution (the {\sc HEALPix} resolution
parameter nside = 2048). In each set, maps are the \emph{lensed} CMB 
temperature plus the {\sc Planck}-HFI nominal Gaussian noise.

\subsection{Full sky lensing reconstruction}

We carry out an integrated potential estimation tool based on the full-sky version of the quadratic 
estimator derived in~\citet{Hu2001o}. We closely follow the prescription given in~\citet{Okamoto2003} 
to build an efficient estimator, so that:
\begin{equation}
\widehat{\phi}_{LM} = \frac{N_L^{\, (0)}}{L(L+1)} \int d\hat{\vec{n}} \,
\, \left( T^{(hp)}(\hat{\vec{n}})\vec{\nabla}T^{(w)}(\hat{\vec{n}}) \right) \, 
\cdot \vec{\nabla}Y_{LM}^*(\hat{\vec{n}}),
\label{eq:philm}
\end{equation}
where $T^{(\mathrm{hp})}$ and $T^{(\mathrm{w})}$ are respectively high-pass filtered and weighted 
lensed CMB temperature field, given by:
%
\begin{eqnarray}
T^{\mathrm{(hp)}}(\hat{\vec{n}}) \, &=& \,  \sum_{lm} \frac{1}{\bar{C}_l^{\mathrm{TT}}} 
\bar{T}_{lm}  Y_{lm}(\hat{\vec{n}})\\\nonumber
T^{\mathrm{(w)}}(\hat{\vec{n}}) \,  &=& \,  \sum_{lm} \frac{C_l^{\mathrm{TT}}}{\bar{C}_l^{\mathrm{TT}}} 
\bar{T}_{lm}  Y_{lm}(\hat{\vec{n}}).
\end{eqnarray}
The covariant derivative operator $\vec{\nabla}$ applied on the
spherical harmonics can be expressed in term of the spin~$\pm 1$ projectors
$e_{\pm} = e_\theta \pm i e_\phi$ and the spin weighted spherical
harmonics, as explained in~\citet{Okamoto2003}.  
The quantities appearing in Eq.~(\ref{eq:philm}) can thus be calculated by 
direct and inverse (spin zero) spherical harmonics and spin-weighted
(spin $\pm 1$) spherical harmonics transforms. As for the estimator
normalisation, quoted $N_L^{\, (0)}$, it identifies with the Gaussian
contribution to the estimator variance; For its expression, we refer to Eq.~(34)
 in~\citet{Okamoto2003}. 
%
\begin{figure*}[t]
  \centering 
    \vfill
        \includegraphics*[angle=90,width=0.47\textwidth]{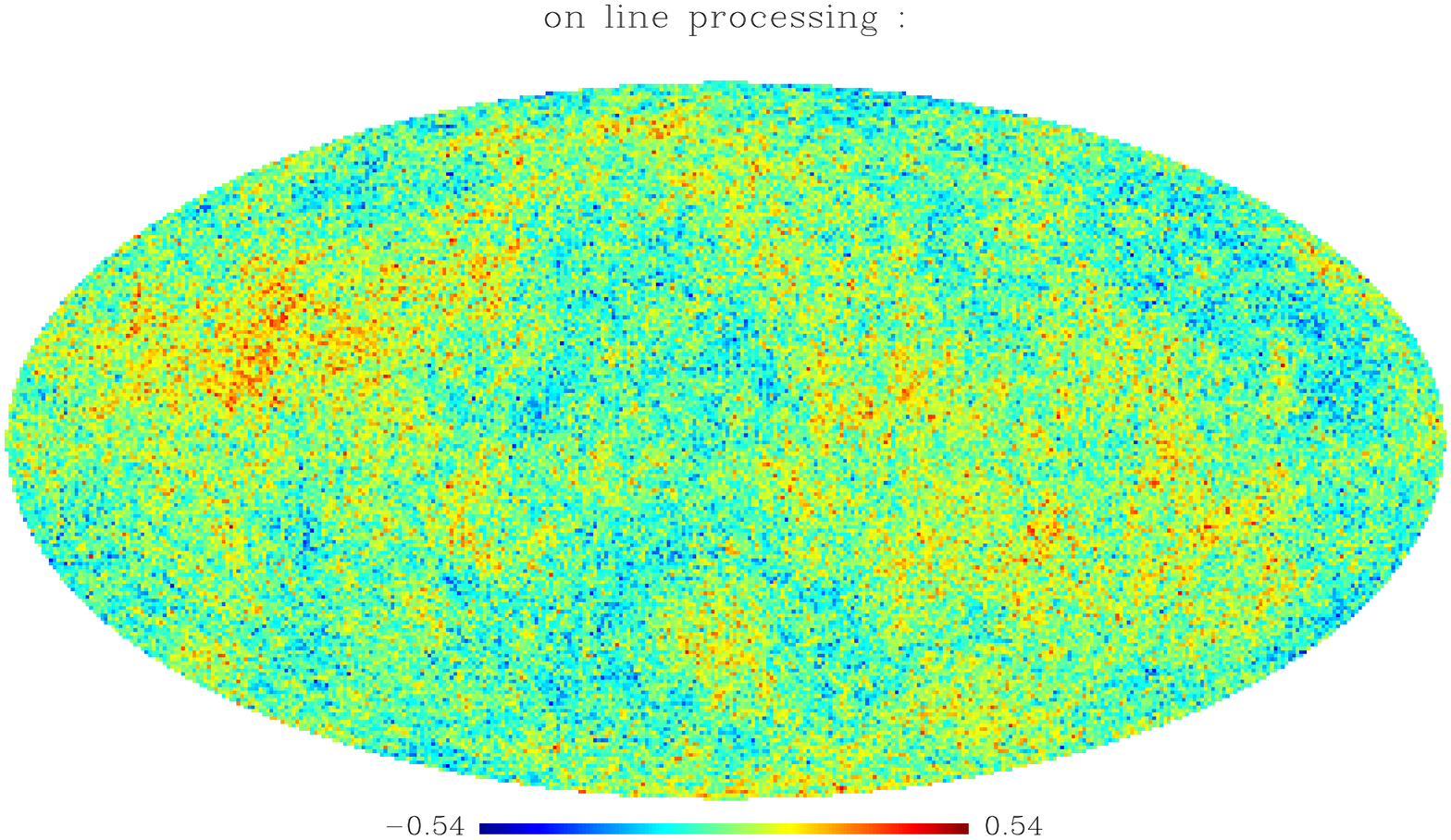} 
     \hfill
        \includegraphics*[angle=90,width=0.47\textwidth]{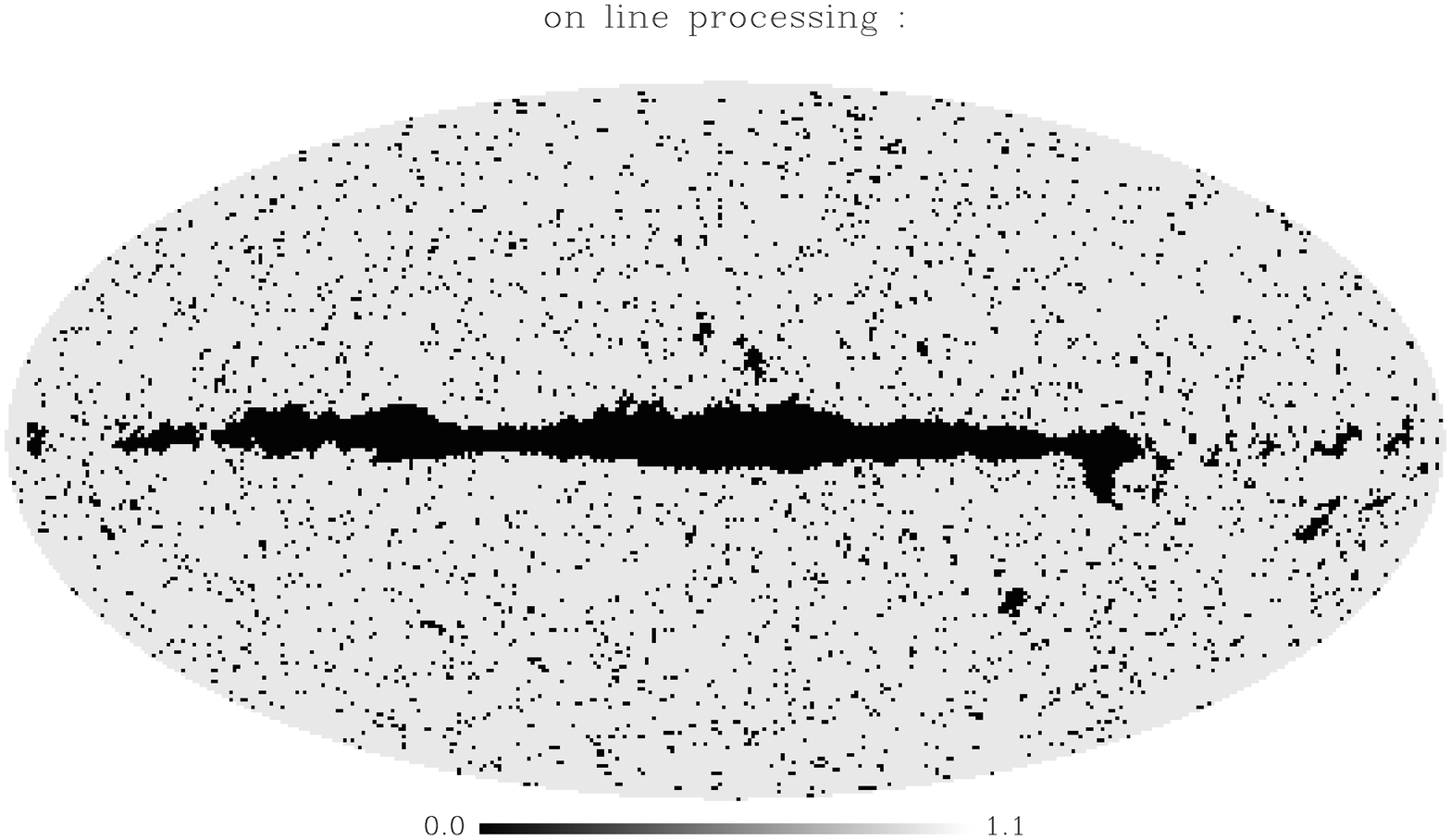} 
     \vfill
        \includegraphics*[angle=90,width=0.47\textwidth]{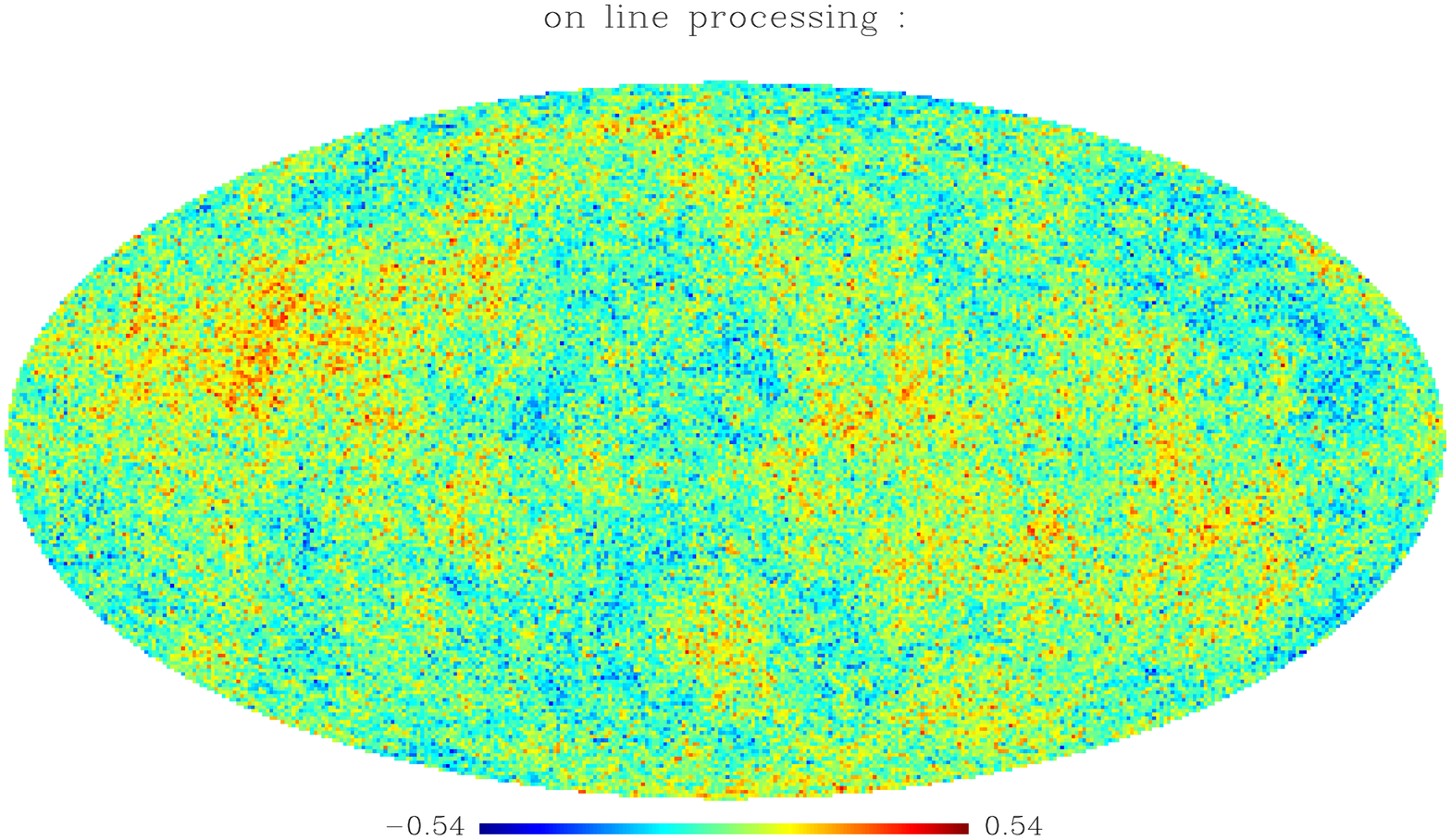} 
     \hfill
        \includegraphics*[angle=90,width=0.47\textwidth]{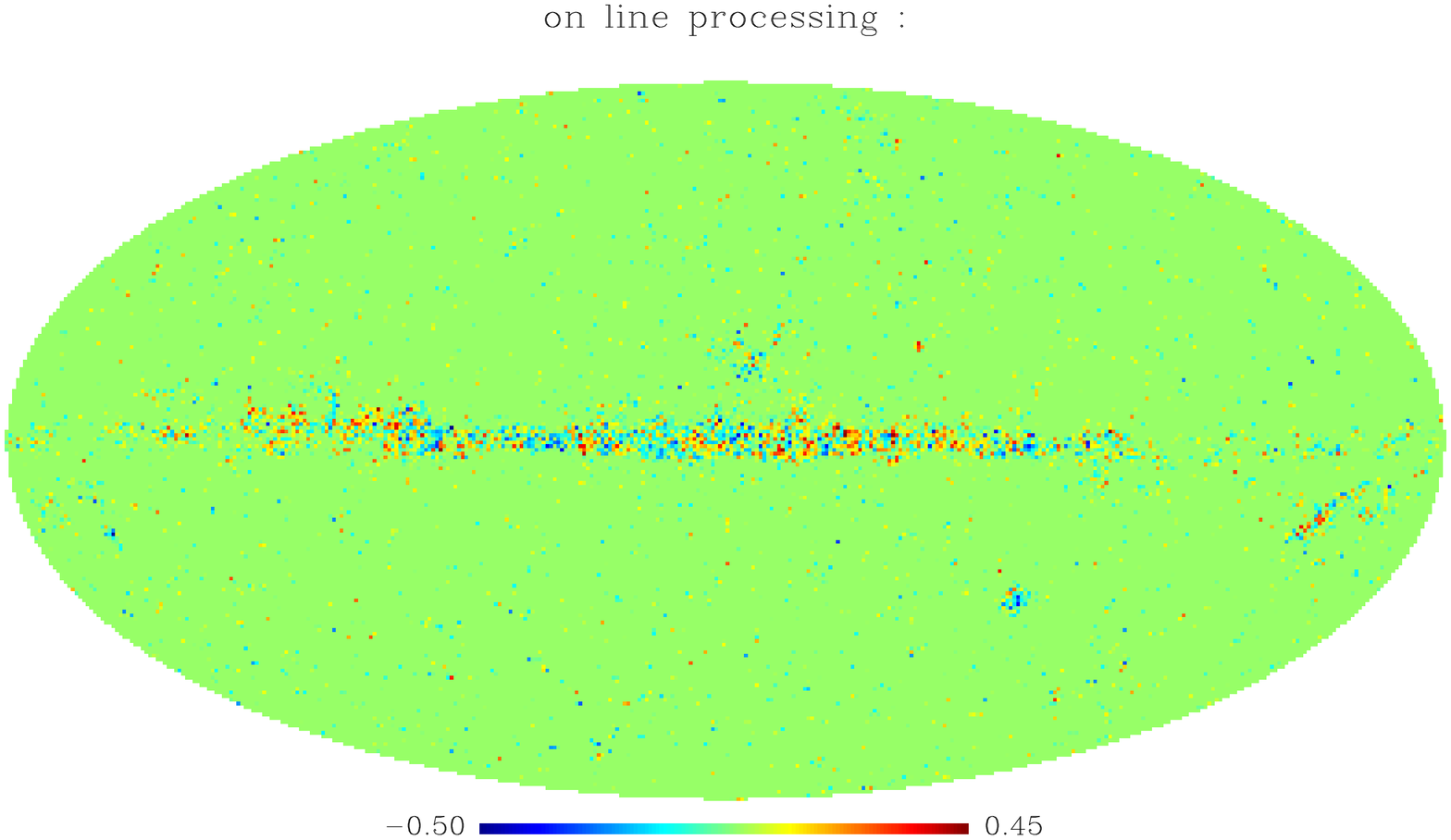}
     \vfill
  \caption{All-sky maps in Galactic Mollweide projection. Upper panels: (left) a {\sc Planck}-HFI synthetic 
CMB temperature map in milliKelvin, (right) the union mask defined in Sect.~\ref{sec:results_inpainting}. The 
grey region shows the observed pixels whereas in black are the rejected ones. Lower panels: left panel shows the restored
 CMB map obtained by applying the \emph{inpainting} process (described in Sect.~\ref{sec:inpainting}) on 
the previous upper-left CMB map masked according to the union mask and the right panel shows the difference between the 
input (upper-left) CMB map and the restored (lower-left) one. }
  \label{fig:mapmask}
\end{figure*}
%
Then, extending to the spherical case the calculations reviewed in Sect.~\ref{sec:formalism},
 the covariance of the integrated potential field estimator $\widehat{\phi}_{LM}$, averaged over
 an ensemble of CMB and gravitational potential fields realisations, depends on the
potential APS, so that:
\begin{equation}
\langle \widehat{\phi}_{LM}^*\widehat{\phi}_{L'M'} \rangle \, =\,  \delta_{LL'} \delta_{MM'}\, 
\left(C_L^{\, \phi\phi} + N_L^{\, (0)} + N_L^{\, (1)}+ N_L^{\, (2)}\right),
\label{eq:spherecldd}
\end{equation}
where $N_L^{\, (0)}$, $N_L^{\, (1)}$ and $N_L^{\, (2)}$ are zeroth, first and
second order in $C_L^{\phi\phi}$ noise terms respectively. To calculate
the first sub-dominant noise term, one can use the expression
derived in the flat-sky approximation by~\citet{Kesden2003}. Likewise, 
the second order term is given in~\citet{Hanson2009}.
 As these terms are an order of magnitude smaller than the dominant $N_L^{\, (0)}$ term and 
because the flat-sky approximation is known to be robust for the potential estimator noise 
calculation~\citep{Okamoto2003}, we assume and will verify that the deviation due to 
the flat-sky approximation is negligible. 

Within the framework of Monte-Carlo analysis, we build a projected potential APS estimator so that:
\begin{equation}
\widehat{C}_L^{\, \phi\phi} \, = \, \frac{1}{N} \sum_{i=1}^{N} \left[
  \frac{1}{2L+1} \sum_{M} |\widehat{\phi}_{LM}^i|^2 \right] - (N_L^{\, (0)} +
N_L^{\, (1)} + N_L^{\, (2)}),
\label{eq:sphereesticldd}
\end{equation}
where $\widehat{\phi}_{LM}^{\, i}$ is the integrated potential field estimate on the $i^{th}$ 
CMB temperature realisation, $i\in\{1,\ldots N\}$. Note that since $N_L^{\, (1)}$ and 
$N_L^{\, (2)}$ depend on the potential APS itself, it should be evaluated and subtracted iteratively. 
Here we calculate it once from the theoretical integrated potential APS.

\begin{figure*}[t]
  \centering
  \vfill
  \includegraphics*[width=0.49\textwidth,height=185pt]{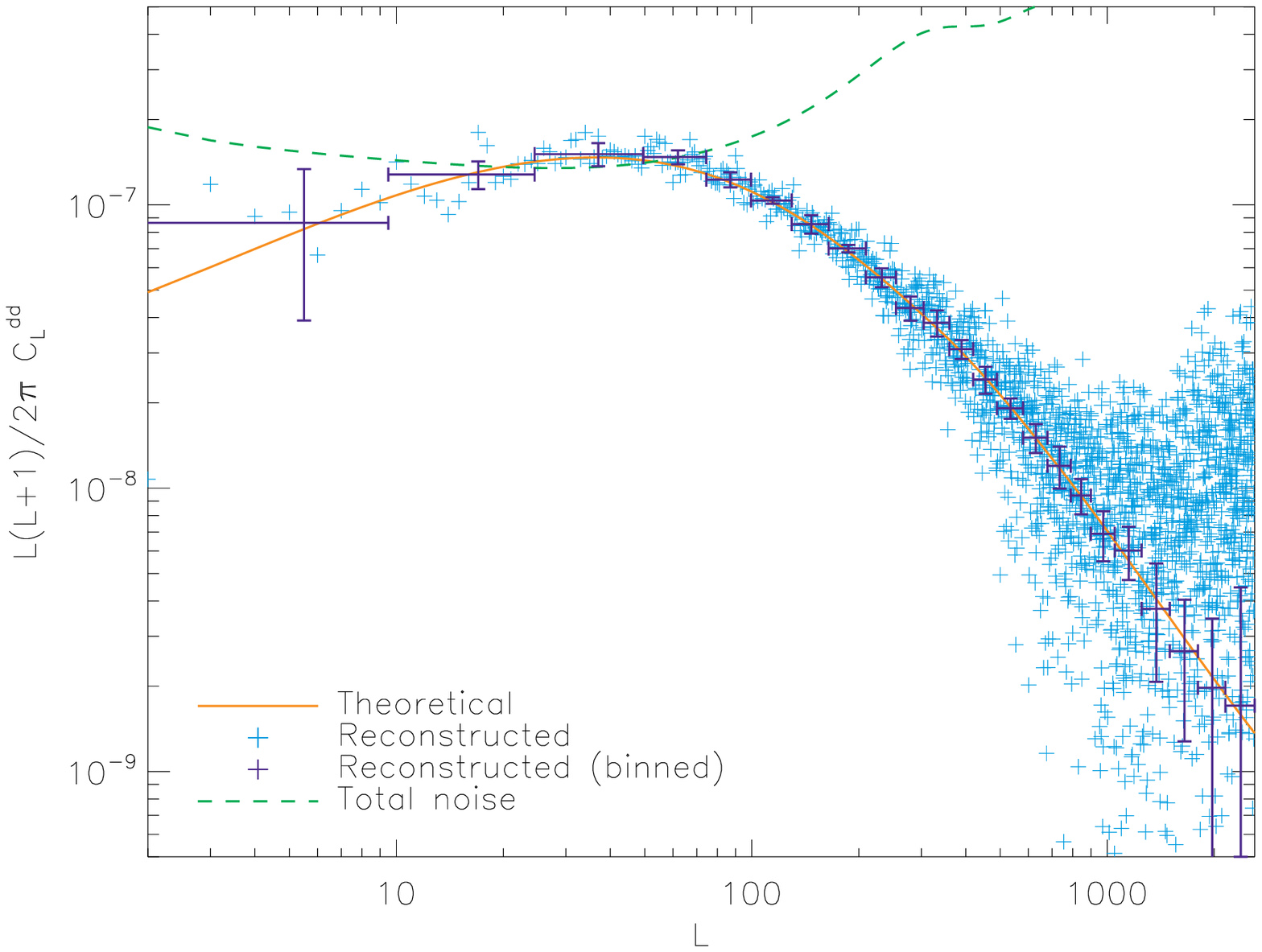}
  \hfill
  \includegraphics*[width=0.49\textwidth,height=185pt]{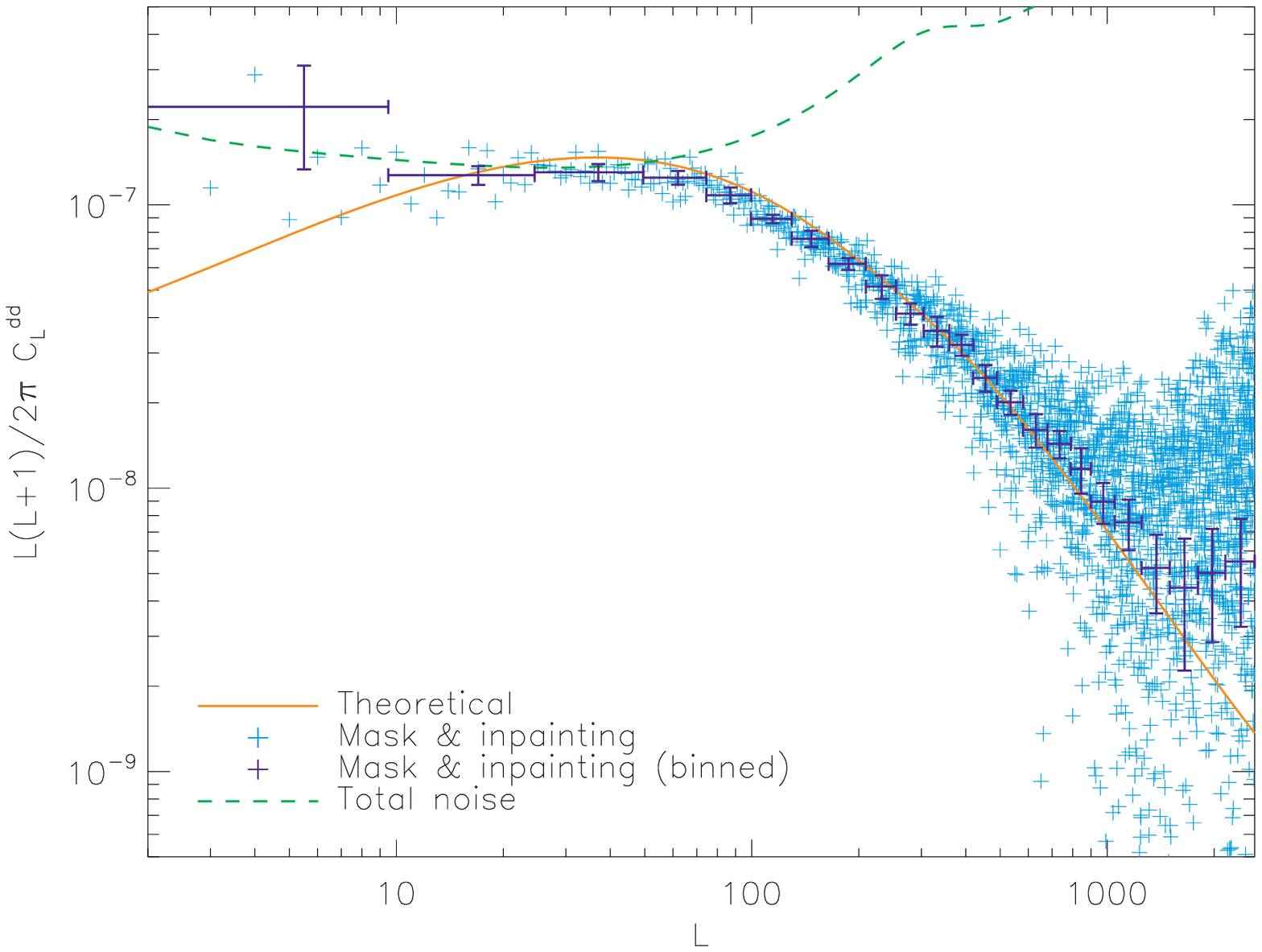}
  \vfill
  \includegraphics*[width=0.49\textwidth,height=160pt]{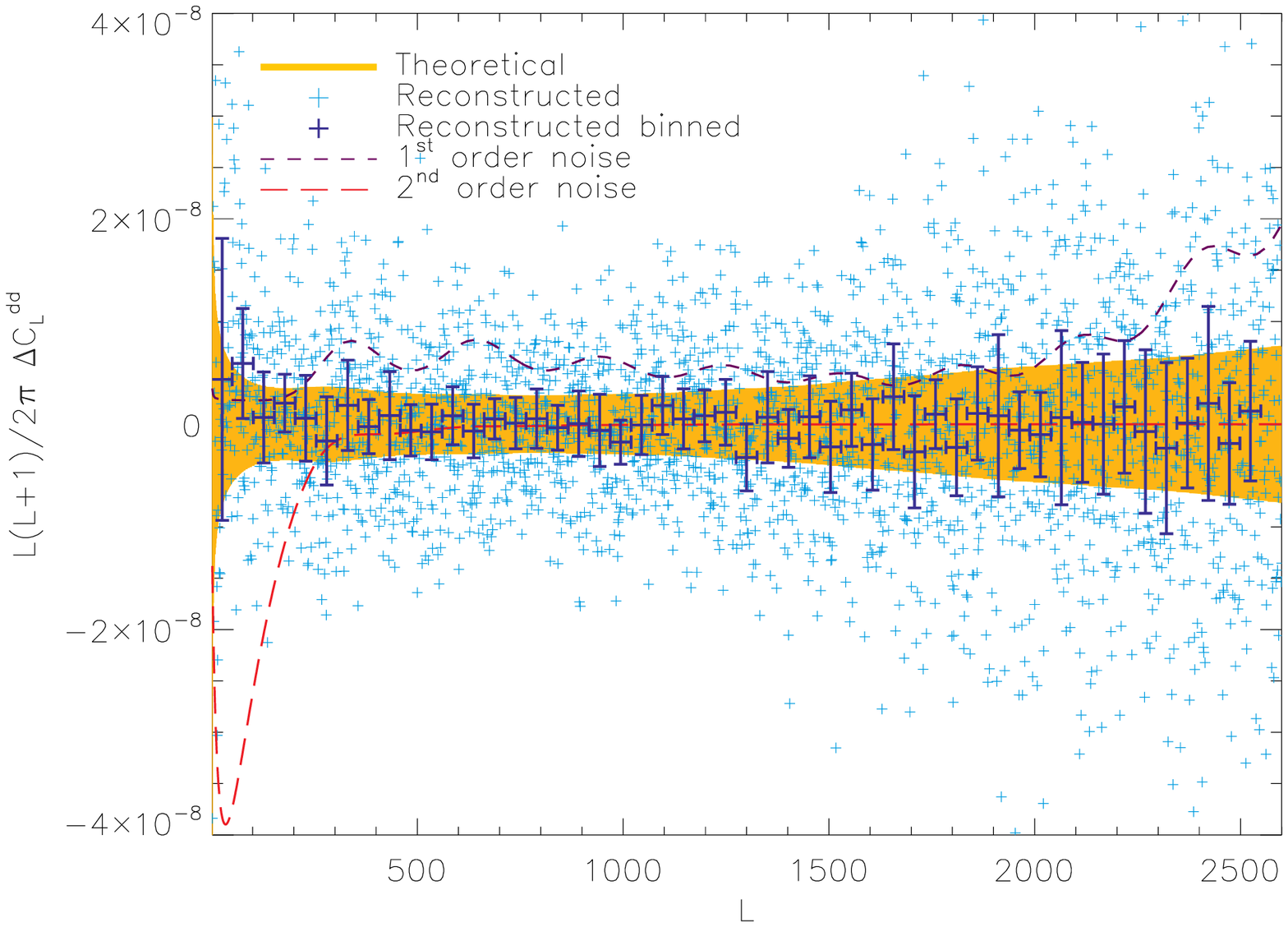}
  \hfill
  \includegraphics*[width=0.49\textwidth,height=160pt]{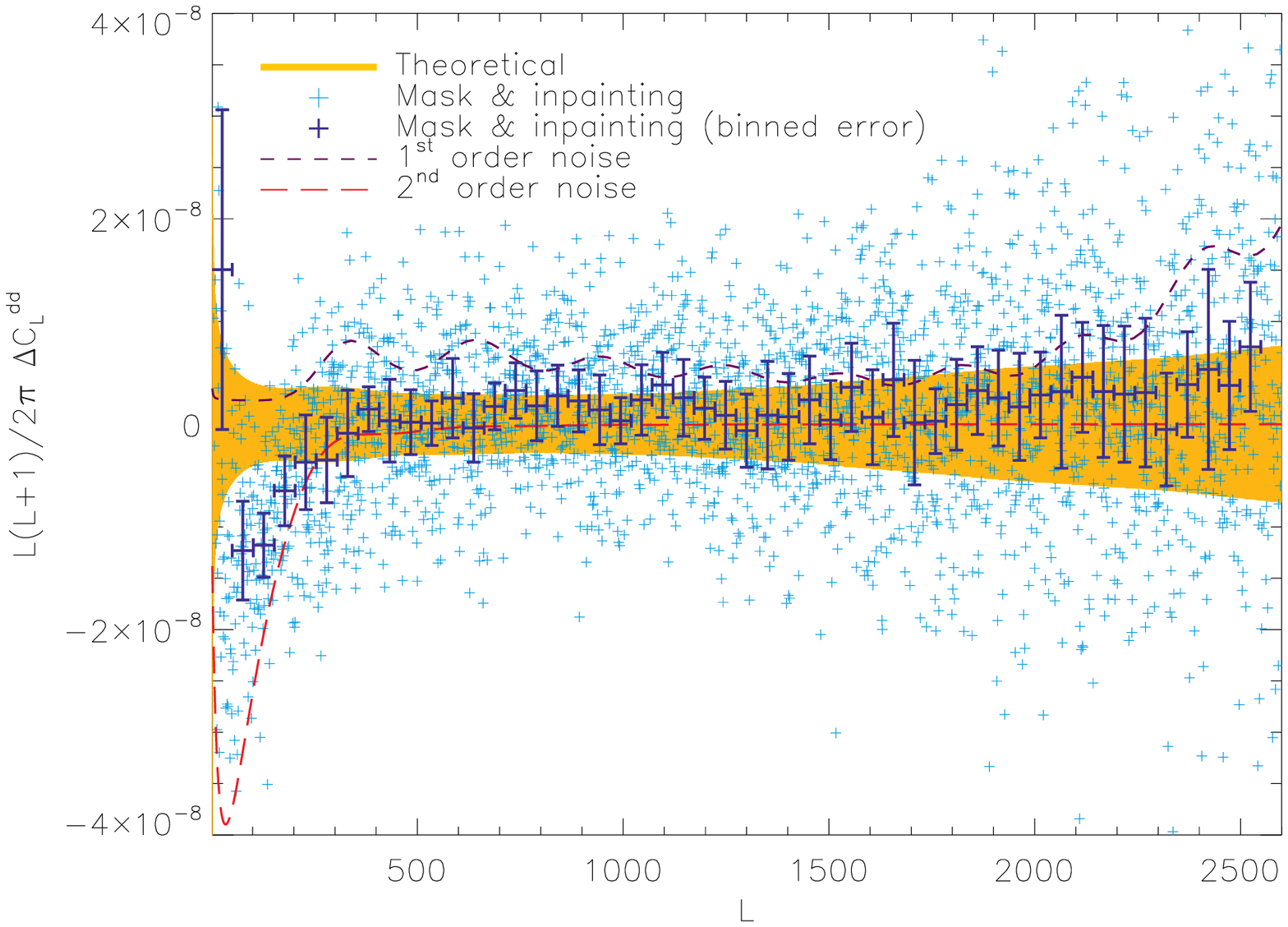}
  \vfill
  \caption{Impact of the masking corrected by an \emph{inpainting} process. The left panels are for the full-sky 
{\sc Planck} synthetic lensed temperature maps whereas the right panels compile the results drawn from the masked 
lensed temperature maps restored with the \emph{inpainting} process described in Sect.~\ref{sec:inpainting}.
\emph{Upper panels}: the reconstructed deflection APS $\widehat{C}_L^{\mathrm{\, dd}}$, obtained from the projected 
potential APS estimate of Eq.~(\ref{eq:sphereesticldd}). The (light blue/grey) crosses show the $\widehat{C}_L^{\mathrm{\, dd}}$
per multipole and the (dark blue/black) data points are the band-power $\widehat{C}_L^{\mathrm{\, dd}}$. The horizontal intervals 
represent the averaging multipole bands and the vertical ones the 1$\sigma$ error. The (orange solid) line figures the fiducial 
deflection angle APS $C_L^{\mathrm{\, dd}}$ and the (green dashed) line the total noise of the quadratic estimator ($N_L^{\, (0)} +$
$N_L^{\, (1)} + N_L^{\, (2)}$). \emph{Lower panels}: the bias of the deflection APS reconstruction $\Delta C_L^{\mathrm{\, dd}}$
defined as the difference between the reconstructed deflection APS $\widehat{C}_L^{\mathrm{\, dd}}$ and the input deflection APS
$C_L^{\mathrm{\, dd}}$. Consistently with the upper panels, (light blue/grey) crosses are $\Delta C_L^{\mathrm{\, dd}}$ per multipole and 
(dark blue/black) data points are the band power $\Delta C_L^{\mathrm{\, dd}}$ with the associated averaging multipole widths 
(horizontal intervals) and 1$\sigma$ error (vertical intervals). The large (orange/grey) band shows the analytical $\pm 1\sigma$
errors per band power expected for the quadratic estimator (see Eq.~\ref{eq:fisher_error}). Finally, the lines show the non-Gaussian
 sub-dominant noise terms at the first-order (violet/dashed) and at the second-order (red/long dashed) in $C_L^{\, \phi\phi}$.
}
  \label{fig:masking}
\end{figure*}

Finally, we test our APS estimator on a set of 10 lensed CMB temperature maps of 50 millions of pixels, 
including the nominal {\sc Planck} noise, generated as described in Sect.~\ref{sec:spheresimu}. As in
 the flat-sky case, sums in the spherical harmonic space are cut at L$_{\rm{max}}$=2600.
The results, compiled in the form of an integrated potential APS estimate averaged over the 10 trials 
(see Eq.~(\ref{eq:sphereesticldd})), are shown in the left panels of Fig.~\ref{fig:masking}.


\subsection{Inpainting the mask}
\label{sec:inpainting}

We choose to take into account the cutting effect of the temperature map before any lensing 
reconstruction rather than making any changes in the quadratic estimator (given in Eq.~\ref{eq:philm})
 to account for the mask. This approach is motivated by the fact that the high quality and the large 
frequency coverage of the {\sc Planck} data allow to reconstruct the CMB temperature map on roughly 
$90\%$ of the sky. It therefore suggests that a method intended to fill the gap in the map can be 
applicable. 

Several of such methods, referred to as \emph{inpainting}, have been recently developed 
since the pioneering work of \citet{Masnou1998}. The general purpose of these methods is to restore
 missing or damaged regions of an image to retrieve as far as possible the original image. For the 
CMB lensing reconstruction, the ideal \emph{inpainting} method would lead a restored map 
having the same statistical properties than the underlying unmasked map. To use a notion
briefly mentioned in the section~\ref{sec:gmca}, the masking effect can be thought of as a loss 
of \emph{sparcity} in the map representation: the information required to define the map has been 
spread across the spherical harmonics basis. That's why, the \emph{inpainting} process can also
be thought of as a restoration of the CMB temperature field \emph{sparcity} in a conveniently chosen 
waveform dictionary. %
%
%
\citet{Elad2005} introduced a sparsity-based 
technique to fill in the missing pixels. This method has been extended to the sphere 
in~\citet{Abrial2008,inpainting:abrial06}. In a nutshell, the masked CMB map is modeled as follows~:
\begin{equation}
\mathcal{T}(\hat{\vec n}) = \mathcal{M}(\hat{\vec n})T(\hat{\vec n})
\end{equation}
where $\mathcal{M}(\hat{\vec n})$ stands for a binary mask the entries of which are one when the pixel 
is observed and zero when it is missing. As emphasized in \citet{Elad2005,Abrial2008}, if 
$T(\hat{\vec n})$ has a sparse representation in a given waveform dictionary $\mathcal{D}$ 
(see \ref{sec:gmca}), masking is likely to degrade the sparsity the CMB map in $\mathcal{D}$.
 Let $\{d^j(\hat{\vec n})\}$ be the set 
of vector that forms the dictionary $\mathcal{D}$. Let $\alpha_{j}$ denote the scalar product
 (so-called coefficients) between 
$T(\hat{\vec n})$ and $d^j(\hat{\vec n})$~: $\alpha_j = \langle T(\hat{\vec n}) , 
d^j(\hat{\vec n}) \rangle$. For the sake of simplicity, we further assume the set 
$\{d^j(\hat{\vec n})\}$ forms an orthonormal basis. Recovering the missing pixel can then be 
made by for a solution that minimizes the sparsity of the ${\mathcal{T}}(\hat{\vec n})$ in $\mathcal{D}$.
 As in \ref{sec:gmca}, an appropriate sparsity estimate of ${\mathcal{T}}(\hat{\vec n})$ in $\mathcal{D}$ 
consists in measuring the sum of the absolute values of $\alpha_j = \langle {\mathcal{T}}(\hat{\vec n}) , 
d^j(\hat{\vec n}) \rangle$. The recovered CMB map is then obtained by solving the following 
optimization problem~:
\begin{equation}
\min_{\{ {\mathcal{T}}(\hat{\vec n}) \}} \sum_j | \langle {\mathcal{T}}(\hat{\vec n}) , d^j(\hat{\vec n}) \rangle 
| \, \mbox{ s.t. } \left \|\mathcal{T}(\hat{\vec n}) - \mathcal{M}(\hat{\vec n})T(\hat{\vec n}) \right \| < \epsilon
\end{equation}
where $\epsilon$ stands for the reconstruction error. It has been shown in~\citet{Abrial2008} that 
this inpainting technique leads to very good CMB recovery results. Our inpainting algorithm can be found 
in~\citet{inpainting:abrial06,Abrial2008} and its implementation is based on the Multi-Resolution 
on the Sphere (MRS) package\footnote{http://jstarck.free.fr/mrs.html}.  
In the following, we seek to assess the impact of the 
\emph{inpainting} mask correction in {\sc Planck}-like maps on a CMB lensing reconstruction.

\subsection{Effect of inpainting}
\label{sec:results_inpainting}

First, we have to choose a realistic mask, which could apply to the forthcoming {\sc Planck} 
temperature map. However, depending on the details of the component separation pipeline the different methods 
developed in the {\sc Planck} consortium \citep[see][]{Leach2008}, yield to slightly different masks. 
Furthermore, the mask size is not a necessary criteria for the final choice of the component 
separation method that will be selected for the {\sc Planck} data analysis. We thus adopt a 
conservative approach, which consists in choosing the union of the masks provided by each of the 
methods at the time of the Component Separation {\sc Planck} \emph{Working Group} second 
challenge~\citep{Leach2008}. Such a mask, hereafter referred to as the \emph{union mask}, rejects 
about 11$\%$ of the sky, as shown in the Fig~\ref{fig:mapmask}.  

Then the 10 {\sc Planck}-like lensed CMB temperature maps we have generated (see Sect.~\ref{sec:spheresimu})
 are masked according to the \emph{union mask} and then restored by applying the \emph{inpainting} 
method described in~\citet{Abrial2008}. From each of these mask corrected maps, we extract a projected 
potential field using the quadratic estimator of Eq.~(\ref{eq:philm}). As previously, the results 
are compiled in the form of the average projected potential APS $\widehat{C}_L^{\, \phi\phi}$ 
following Eq.~(\ref{eq:sphereesticldd}). The reconstructed deflection APS, 
given by $\widehat{C}_L^{\mathrm{\, dd}} = L(L+1)\widehat{C}_L^{\, \phi\phi}$, as well as the bias between 
estimated and fiducial deflection APS $\Delta C_L^{\mathrm{\, dd}}$ are shown in the right panels of 
Fig.~\ref{fig:masking}. 

We find that the mask corrected by the inpainting results in a marginal increase ($\sim 4 \%$) of the 1$\sigma$
 errors on the estimated deflection APS (hence on the projected potential APS). Masking and inpainting causes 
an increase of the reconstructed APS bias $\Delta C_L^{\mathrm{\, dd}}$ arising mostly at 
large angular scale corresponding to multipole $L<300$. However, this bias is weaker than the 
sub-dominant second-order in $C_L^{\, \phi\phi}$ non-Gaussian bias. Fig.~\ref{fig:masking} shows a clear 
increase of power in the very first multipole band ($2<l<10$). In this multipole range, {\sc Planck}
 is not expected to achieve a good reconstruction of the potential APS~\citep{Hu2002}. From the multipole 
$L=300$ up to $L=2600$, the bias stays below the first-order in $C_l^{\phi\phi}$ non-Gaussian bias and is 
compatible with the theoretical $1\sigma$ errors expected for the quadratic estimator. From fully 
controlling the inpainting impact, one might want to push further the study by analytically calculating
 or Monte-Carlo estimating the mask induced bias. However, it is not mandatory for 
reconstructing the projected potential APS with {\sc Planck}. The masking effect, once corrected by 
inpainting, becomes a sub-dominant systematic effect that can be safety neglected.

\begin{figure}[t]
  \centering
  \vfill
  \includegraphics*[width=245pt,height=185pt]{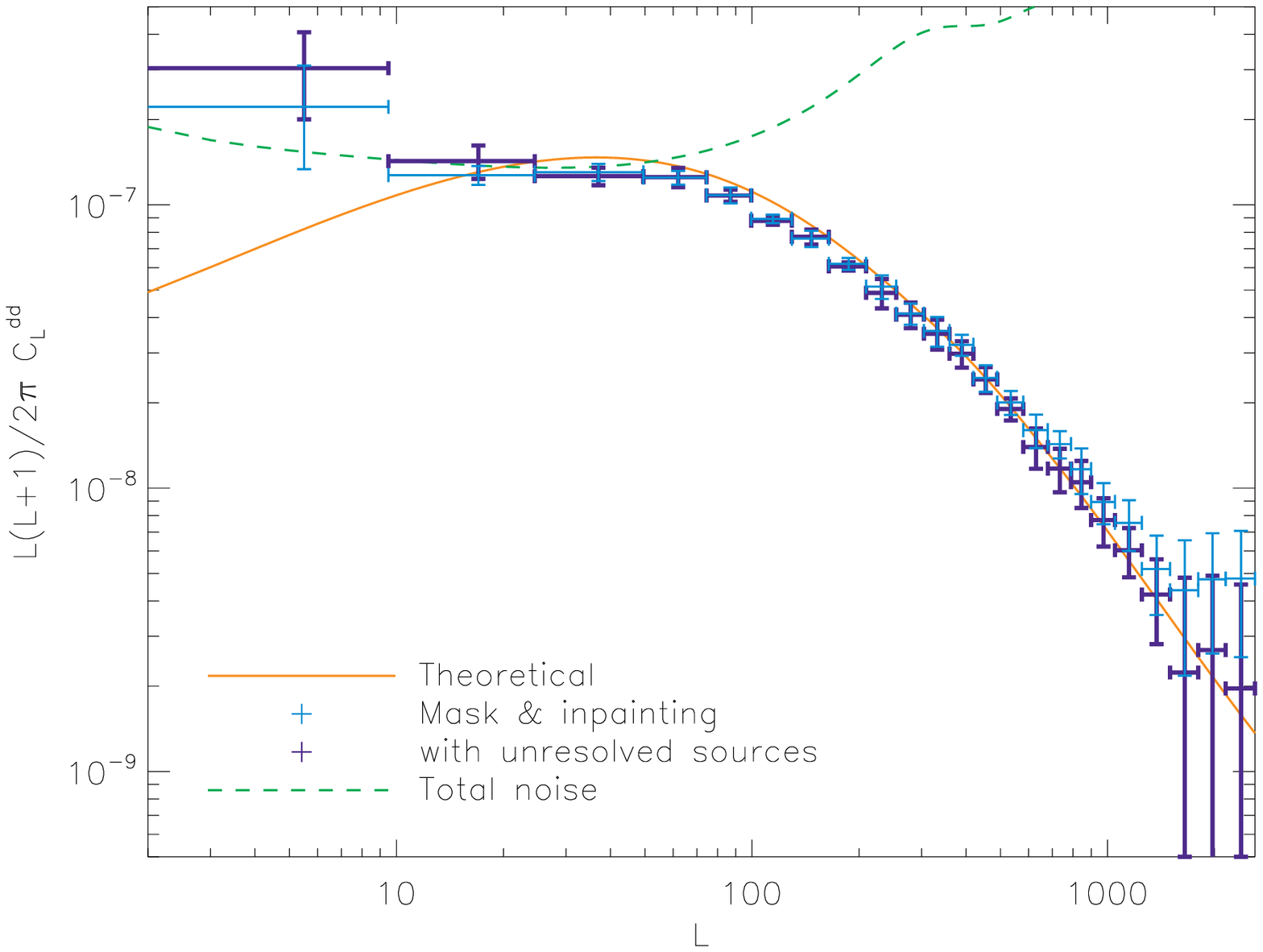}
  \vfill
  \includegraphics*[width=245pt,height=160pt]{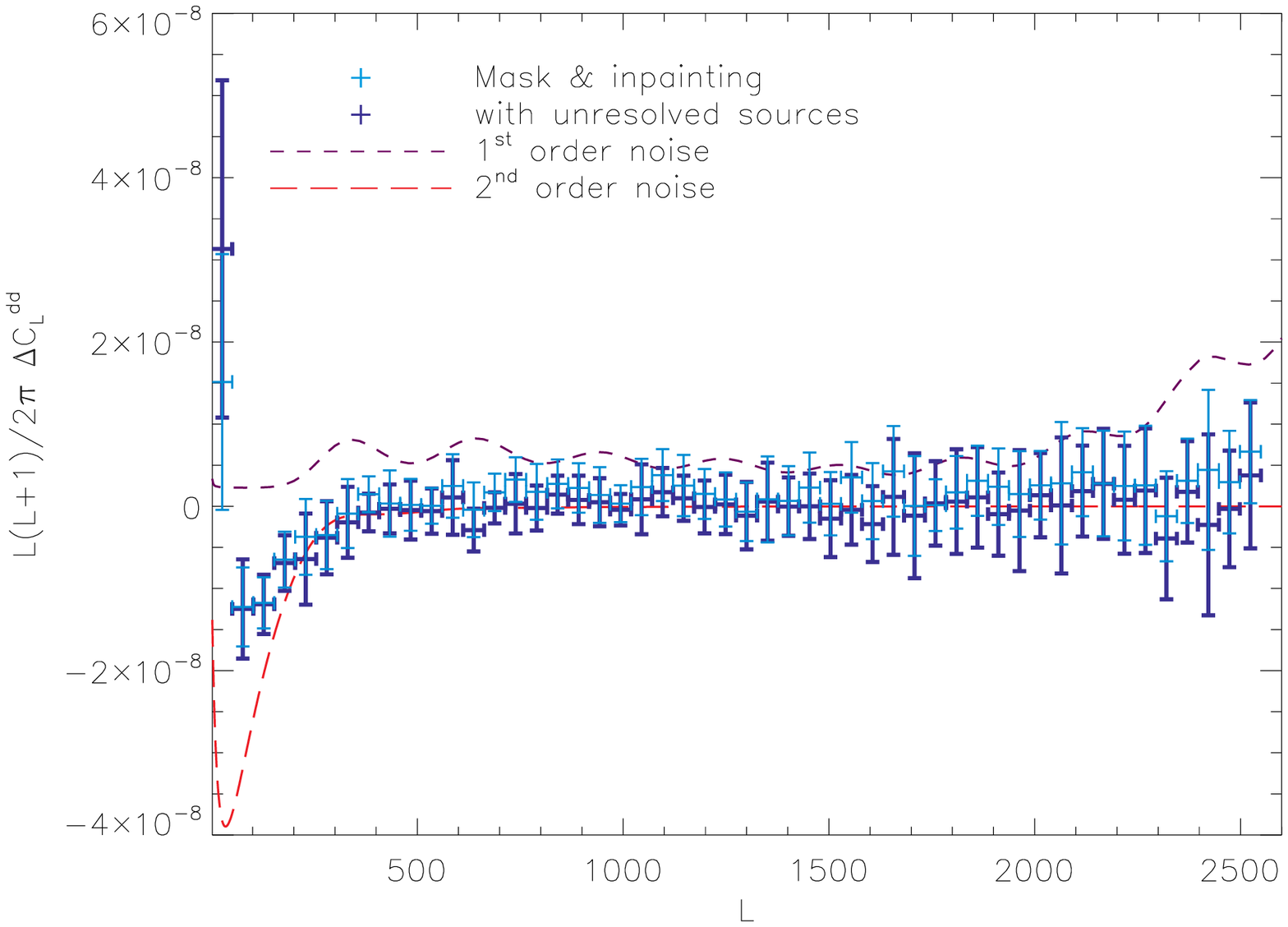}
  \vfill
  \caption{\modif{Robustness of the inpainting to the unresolved point sources. The upper panel shows the reconstructed deflection 
APS whereas the lower panel the bias on the deflection APS reconstruction. The (dark blue/black) data points show the result 
of the full-sky quadratic estimation on the set of lensed CMB maps with point sources residuals, which have been masked then 
restored with the inpainting technique, as described in Sect.~\ref{sec:results_pointsources}. For comparison, the results 
obtained in Sect.~\ref{sec:results_inpainting}, in the case without any sources residuals, have been copied out on here as 
(light blue/grey) crosses. The fiducial analytical deflection APS as well as the total reconstruction noise and the two APS 
bias are shown following the same representation code as in Fig.~\ref{fig:masking}, and horizontal and vertical intervals 
have the same meaning as described in the Fig.~\ref{fig:masking} caption.  For a sake of readibility, only the 
\emph{band power} reconstructed APS and APS bias are represented, and the theoretical error bars by multipole bins are not shown.}
 }
  \label{fig:pointsources}
\end{figure}

\modif{ 
\subsection{Robustness against the unresolved point sources}
\label{sec:results_pointsources}

 Up to now, we have handled independently two important issues linked to the presence of foreground emissions 
in the observation maps, the impact of the foreground residuals after component separation with GMCA in 
Sect.~\ref{sec:flatestimation} and the impact of the masking corrected with the inpainting method in the 
previous subsection (Sect.~\ref{sec:results_inpainting}). We found that none of them compromises our ability 
to reconstruct the deflection APS. In a more realistic approach, these two issues should be handled altogether,
 as the inpainting process is intended to be applied on a CMB map contaminated by foreground residuals. The presence
 of foreground residuals is susceptible to harden the inpainting process and consequently degrade the CMB 
lensing recovery. Here, we assess the robustness of the deflection reconstruction on masked and inpainted CMB
 maps when adding infra-red point sources residuals. This choice is motivated for two reasons. The point sources 
residuals after component separation is a well-known matter of concern in any CMB non-gaussianities analysis 
and the emission of the infra-red sources population is one of the major foreground contaminant at the 
{\sc Planck}-HFI observation channels.

We use the full-sky map of infra-red point sources residuals after a component separation using GMCA, we had 
estimated in Sect.~\ref{sec:gmca}. This point sources residuals is added to the 10 synthetic {\sc Planck} 
lensed CMB temperature maps described in Sect.~\ref{sec:spheresimu}. Then we redo the same analysis than 
previously in Sect.~\ref{sec:results_inpainting}: the \emph{union mask} is applied to the maps, cutting out 
the brightest infra-red sources, which have been detected during the Component Separation {\sc Planck} 
\emph{Working Group} second challenge~\citep{Leach2008}. The 10 masked maps are restored using the 
\emph{inpainting} method before being ingested in the full-sky quadratic estimator of the projected potential 
field. The results of the whole analysis are presented in the form of the average reconstructed deflection 
APS and bias, and shown in Fig.~\ref{fig:pointsources}.     

We find that the inpainting performances are only marginally degraded (at $\leq 1\sigma$ level) by the presence
of point sources residuals within the CMB maps, and this degradation occurs mainly at the two multipole 
extremes. At the lower multipoles ($L<30$), the APS deflection reconstruction suffers from a $1\sigma$ 
increase of the bias, whereas at the higher multipoles, only the error bars increase. 
We conclude that the inpainting method succeeds in keeping the statistical properties of the CMB map unchanged even in 
presence of highly non-Gaussian foreground residuals and it is a qualified method to handle the masking issue when 
seeking at a CMB lensing recovery. In addition, the results compiled in the Fig.~\ref{fig:pointsources} give
the total impact of point sources on the deflection reconstruction, as they account for both the masking
of the bright detected sources and the unresolved residuals. We report that point sources are responsible for
a total $13\%$ increase of the $1\sigma$ errors on the reconstructed APS deflection, mainly induced by the 
unresolved residuals. As a summary, the major nuisance of point sources is related to the masking of the 
bright ones, which tend to increase the bias on the reconstructed deflection APS, whereas the unresolved 
residuals results mainly in an increase of the errors on the deflection retrieval. 
}

%
%

%
%
%
%
%
%
%
\section*{Conclusions}

The High Frequency Instrument (HFI) of the {\sc Planck} satellite, \modif{which has been launched on the 
14$^{\mathrm{th}}$ of May 2009}, has the sensitivity and the angular resolution required to allow  
a \emph{reconstruction} of the CMB lensing using the temperature anisotropies map alone. The pioneer 
works to put evidence of the CMB lensing within the WMAP data are not 
directly applicable or not well-optimized to the {\sc Planck} data. First, one might want to take benefit 
of the efficient component separation algorithms developed for {\sc Planck} before applying a CMB lensing 
estimator rather than to correct the lensing reconstruction from the bias due to the foreground 
emission afterward. Second, we need an efficient and manageable method to take into account the sky cutting
 within the 50 Mega-pixels maps provided by {\sc Planck}. In addition, the CMB lensing is related to 
another burning thematic: characterizing the non-Gaussianities of the temperature anisotropies 
(primordial non-Gaussianities, cosmic string, etc.).  

We have implemented both the flat-sky and the all-sky versions of the quadratic estimator of the 
projected potential field described in~\citet{Hu2001a,Okamoto2003} to apply them on {\sc Planck} 
synthetic temperature maps. First, within the flat-sky approximation, we have prepared a 
\emph{demonstration model}, which consists in running GMCA, a component separation method described 
in~\citet{Bobin2008}, on {\sc Planck} frequency channel synthetics maps, containing the lensed
CMB temperature, the {\sc Planck} nominal instrumental effects (modeled by a white Gaussian noise
and a Gaussian beam) and \modif{the three dominant foreground emissions at the {\sc Planck}-HFI 
observation frequencies, namely the SZ effect, the galactic dust and the infra-red point sources.}
 We have performed a Monte-Carlo analysis to quantify the impact of the foreground residuals after 
the GMCA on the projected potential field and APS reconstructions. Then, we have moved on to the full-sky 
case, using the LensPix algorithm~\citep{Lewis2005} to generate lensed CMB temperature 
maps at the {\sc Planck} resolution. We have performed a Monte-Carlo analysis to tackle 
the masking issue; we have used the \emph{inpainting} method described in~\citet{Abrial2008} to restore 
the {\sc Planck} synthetic temperature maps, masked according to a realistic cut out of 11$\%$ of the sky, 
\modif{ accounting for the bright detected point sources}. By applying the projected potential quadratic 
estimator on these restored maps, we have studied the impact of the inpainting of the mask on the 
{\sc Planck} sensitivity to the projected potential APS. \modif{Finally, we have assessed the total impact
 of the point sources emission, in confronting the inpainting method to the unresolved point sources 
residuals.}      

\paragraph{Results}
\begin{enumerate}
      \item Within our flat-sky \emph{demonstration model}, we found that the reconstruction of the 
projected potential field is still feasible after a component separation using GMCA. More 
quantitatively, the foreground residuals in the GMCA output CMB maps \modif{lead to a $10\%$ increase} 
of the 1$\sigma$ errors on the projected potential APS reconstruction when applying the quadratic 
estimator. The GMCA process results in an increase of the dispersion of the projected potential APS 
reconstruction, but this dispersion remains within the theoretical 1$\sigma$ errors \modif{at all 
angular scales but the $L<60$ multipoles, in which the flat-sky analysis is expected to show some 
limitations anyway}. Such a study dealing with the impact of a component separation process on the CMB 
lensing reconstruction had never been performed before. Our results allow us to assess that applying 
a component separation algorithm on the frequency channel CMB maps before any lensing estimation is 
a well-adapted strategy for the 
sake of the projected potential reconstruction within {\sc Planck}.   
      \item For the full-sky reconstruction of the projected potential APS with {\sc Planck}, we 
report that a realistic 11$\%$ of the sky mask, applied on some {\sc Planck}-nominal lensed CMB 
temperature maps, has a negligible impact on the CMB lensing signal retrieval process, whenever 
it has been corrected by the \emph{inpainting} method of \citet{Abrial2008} beforehand. More 
precisely, the bias on the estimated projected potential APS induced by the mask after inpainting is
always either compatible with the theoretical $1\sigma$ errors (from $l=300$ up to $l=2600$) or 
weaker than the second-order in $C_l^{\phi\phi}$ non-Gaussian bias (in the $l<300$ range). The major 
impact of the inpainting correction on the projected potential APS arises at the larger angular 
scales ($2<l<10$), which are not expected to be well-reconstructed with {\sc Planck}. \modif{In addition,
these results have not significantly changed after the introduction of unresolved point sources residuals. 
When treating the point sources emission in a comprehensive way, we report a $13\%$ increase of the 
$1\sigma$ errors on the reconstructed deflection APS on average, resulting mainly from the unresolved 
point sources residuals, whereas the level of bias is marginally increased at low multipoles}. We conclude 
that applying the inpainting method of \citet{Abrial2008} beforehand is a good strategy to take into account
 the masking issue when seeking at reconstructing the projected potential with {\sc Planck}.   
   \end{enumerate}

\paragraph{Perspectives} 
Our results on the CMB lensing reconstruction are the first step to elaborate a complete
 analysis chain dedicated to the projected potential APS reconstruction with the {\sc Planck} data. 
Such a CMB lensing reconstruction pipeline should involve a \modif{component 
separation and a bright point sources detection} followed by an algorithm to correct from the mask 
(e.g. the inpainting method) before applying a quadratic estimator of the projected potential field 
on the resulting CMB temperature map.   

We plan to go on developing in parallel both the flat-sky and the full-sky reconstruction tools. The 
flat-sky tools will allow us to perform a multi-patches CMB lensing reconstruction in cutting several
hundred patches out of the most foreground cleaned region of the full-sky map. Using such a method 
requires a quantitative study of the impact of the sphere-to-plan projection and the
sharp edge cuts effects beforehand. As a first task, we should test whether our results concerning the 
feasibility of reconstructing the CMB lensing after a component separation and after an inpainting of the mask 
still hold when dealing with the fully realistic {\sc Planck} simulation (including e.g. non axisymmetric 
beam, inhomogeneous noise and correlated foreground emissions). As long as we can demonstrate we have a sufficient 
control on systematics, we will be ready to measure the projected potential APS with {\sc Planck} alone. 
Such an additional cosmological observable is expected to enlarge the investigation field accessible to 
the {\sc Planck} mission from the primordial Universe to us.

%
%
%

\begin{acknowledgements}
We warmly thank Duncan Hanson for providing us the second-order non-Gaussian noise term 
biasing the projected potential APS estimation and for helpful discussions. We also would like 
to thank Martin Reinecke for his help on the HEALPix package use. We acknowledge use 
of the CAMB, LENSPix, HEALPix-Cxx and MRS packages. 
This work was partially supported by the French National Agency for Research (ANR-05-BLAN-0289-01 and ANR-08-EMER-009-01).
\end{acknowledgements}

%
%
%
%
\bibliographystyle{bibtex/aa}
\bibliography{Lensing}

\end{document}